\documentclass{aa}

\usepackage{graphicx}
\usepackage{amsmath}
\usepackage{amssymb}
\usepackage{txfonts}

\newcommand{\Teff}{\mbox{$T_{\mbox{\rm \tiny eff}}$}}

\begin{document}

\title{Near-IR spectroscopic ages of massive star clusters in M82.}

\author{A. Lan\c{c}on\inst{1}
   \and
        J.S. Gallagher, III\inst{2}
   \and
        M. Mouhcine\inst{3}
   \and
        L.J. Smith\inst{4,5}
   \and
        D. Ladjal\inst{6}
   \and
        R. de Grijs\inst{7,8}
       }

\offprints{A. Lan\c{c}on \email{lancon@astro.u-strasbg.fr}  }

\institute{
 Observatoire Astronomique de Strasbourg,
 Universit\'e L.\,Pasteur \& CNRS (UMR 7550), Strasbourg, France
\and
 Department of Astronomy, University of Wisconsin-Madison, WI, USA
\and
 Astrophysics Research Institute, Liverpool John Moores University,
 UK
\and
 Space Telescope Science Institute and European Space Agency, Baltimore, MD, USA
\and
 Department of Physics and Astronomy, University College London, London, UK
\and
 Institute of Astronomy, Katholieke Universiteit, 
 Leuven, Belgium
\and
 Department of Physics \& Astronomy, The University of Sheffield,
 Sheffield, UK
\and
 National Astronomical Observatories, Chinese Academy of Sciences,
 Beijing, China
}

\date{Received 20/01/08 - Accepted 10/04/08}

\authorrunning{A. Lan\c{c}on et al.}

\abstract
{Like other starburst galaxies, M\,82
hosts compact, massive ($>5\times 10^5$\,M$_{\odot}$) 
young star clusters that are interesting both in their
own right and as benchmarks for population synthesis models.
}
{In addition to assessing or reassessing the properties of some of the
brightest near-IR sources in M\,82, this paper addresses the following 
questions. Can population synthesis models at 
$\lambda/\delta \lambda \simeq 750$ adequately reproduce
the near-IR spectral features and the energy distribution of these
clusters between 0.8 and 2.4\,$\mu$m? How do the derived cluster properties
compare with previous results from optical studies? 
}
{
We analyse the spectra of 5 massive clusters in M\,82, using data
acquired with the spectrograph SpeX on the InfraRed Telescope
Facility (NASA/IRTF) and a new population synthesis tool
with a highly improved near-IR extension, based on a recent collection of
empirical and theoretical spectra of red supergiant stars.
}
{We obtain excellent fits across the near-IR with models at quasi-solar
metallicity and a solar neighbourhood extinction law. 
Spectroscopy breaks a strong degeneracy between age and extinction
in the near-IR colours in the red supergiant-dominated phase of evolution.
The estimated near-IR ages cluster between 9 and 30\,Myr, i.e. the ages
at which the molecular bands due to luminous red supergiants are 
strongest in the current models. They do not always agree with optical
spectroscopic ages. Adding optical data sometimes leads to the rejection
of the solar neighbourhood extinction law. This is not surprising 
considering small-scale structure around the clusters, but it has no 
significant effect on the near-IR based spectroscopic ages.
}
{
The observed IR-bright clusters are part of the most recent episode of 
extended star formation in M\,82. The near-IR study of clusters that
are too faint for optical observation adds important elements to
the age distribution of massive clusters in dusty starbursts.
Further joint optical and near-IR spectroscopic studies will provide strong 
constraints on the uncertain physics of massive stars on which population 
synthesis models rest. 
}


\maketitle


\section{Introduction}

Starburst galaxies host large populations of star clusters, and it has
become increasingly clear that global star formation processes
and cluster formation processes are intimately linked (Meurer et al. 1995,
Larsen 2006, Elmegreen 2006). 
This is an important motivation for detailed studies
of young star clusters, their distributions in terms of mass and age,
their mass-to-light ratios and their stellar initial mass functions.

Because star formation originates in molecular clouds, starburst
galaxies tend to be dusty, and young star clusters can be severely
obscured. As beautifully illustrated by near-IR and 
optical Hubble Space Telescope (HST) images
of galaxies such as M\,82 or the Antennae
(de Grijs et al. 2001, Alonso-Herrero et al. 2003,
McCrady et al. 2003, Rossa et al. 2007), the brightest near-IR star clusters
can go undetected at optical wavelengths, and vice-versa. Optical
studies tend to focus on lines of sight through holes in the dust, 
and the derived distributions of cluster properties are subject to biases.
In order to obtain a more complete picture of the cluster populations,
it seems crucial to combine optical and near-IR data.
This can be done with multiband photometry or with spectroscopy.

The importance of near-IR spectroscopy in studies of dusty
star forming galaxies has long been recognised (Rieke \& Lebofsky 1979),
and spectra in the H or K windows have been used to constrain star formation
histories (e.g. Rieke et al. 1993, Oliva et al. 1995,
Lan\c{c}on \& Rocca-Volmerange 1996,
Satyapal et al. 1997, F\"orster Schreiber et al. 2001).

However, near-IR studies of young stellar populations remain difficult. Among
the relevant problems are the following (see main text for details and
references).\\
(i) Evolutionary
tracks for massive stars are uncertain, in particular in the red supergiant
phase which is the predominant source of near-IR light at  ages of
order 10$^7$\,yr. \\
(ii) Model atmospheres for luminous red supergiants are not yet
reliable, and this affects the mapping of isochrones into colour
diagrams.  \\
(iii) The stellar population of a cluster is stochastic,
and the intrinsically small number of bright stars leads to potentially large
spreads and systematics in the expected cluster properties. \\
(iv) The subtraction of very irregular starburst galaxy backgrounds
can be hazardous. \\
(v) It matters to photometric studies
that the shape of the dust attenuation law depends on
the unknown spatial distribution of the dust as well as on the nature
of the dust; there are degeneracies in colours, for instance between age
and extinction. \\
As a result of some or all of the above, 
discrepancies between photometric studies that
do or do not include near-IR passbands can occur.
The consistency between results obtained
from near-IR and from optical spectroscopic data needs to be tested.

With {\em extended} spectroscopy of massive star clusters, by which
we mean spectroscopy that extends through optical
and near-IR wavelengths, it becomes possible to isolate some of the
above aspects.  By using stellar absorption features
exclusively, dust attenuation issues are mostly eliminated. By focusing 
on very massive star clusters, stochasticity effects can be
kept below acceptable levels. Contamination by other
stellar populations on neighbouring lines of sight remains an issue 
in crowded areas, but can be addressed with multicomponent models,
more safely with spectroscopy than with photometry. 
In the end, the spectroscopic
analysis is mostly sensitive to the ingredients of the 
population synthesis models, such as the stellar evolution tracks
and the stellar spectra assigned to each point along theses tracks.
One may expect two types of results. If the best-fit models based
on the near-IR data are not consistent with the optical data, this
points to inadequacies of the input physics of the models, i.e.
likely systematic errors in both optical and near-IR cluster ages. 
Conversely, a good fit to the optical and near-IR data set 
will provide significantly more robust cluster properties
than a fit to a single spectral window or to photometric data.

In this paper, we analyse extended spectra of five 
massive clusters in the nearby starburst galaxy \object{M82}, using 
a population synthesis tool with a highly improved near-IR
spectroscopic extension.  We show that it is possible
to obtain a good representation of all the near-IR features 
(0.8-2.4\,$\mu$m, $\lambda/\delta \lambda \sim 750$). We
investigate whether the near-IR results are consistent with
the optical constraints, and what we can learn about extinction and
the possible origins of difficulties faced in photometric age-dating 
procedures. 

Section \ref{models.sec} introduces the new population synthesis tool.
In Sections \ref{clusters.sec} and \ref{analysis.sec} we describe and
analyse the spectra of star clusters in M\,82. Selected aspects of the 
analysis as well as the issue of stochastic fluctuations are discussed
in Sect.\,\ref{discussion.sec}. Conclusions are summarised in 
Sect.\,\ref{conclusion.sec}.

\section{Synthetic near-IR spectra of young stellar populations}
\label{models.sec}

Multi-wavelength spectroscopic studies of young stellar populations 
have been limited by the absence of complete stellar spectral
libraries with adequate near-IR coverage and resolution.
A reasonable coverage of the HR diagram was available in the libraries
of Pickles (1998) and Lejeune et al. (1998). Both have insufficient
spectral resolution for our current purposes and lack some of the most 
important molecular features of cool stars, in particular 
between 0.9 and 1.5\,$\mu$m. Libraries at higher spectral resolution
are usually limited to selected near-IR spectral windows and their
continua have sometimes been normalised (Meyer et al. 1999,
Wallace et al. 2000, Cenarro et al. 2001, Ivanov et al. 2004 and 
further references therein).
Unfortunately, many of the near-IR features of interest for 
population diagnostics are broad molecular bands that can not be
studied reliably when the spectra in the traditional windows of 
ground based observations are acquired separately (unless high
quality absolute flux calibration is achieved). 

An exception has been the collection
of spectra of cool stars of Lan\c{c}on \& Wood (2000), which covers
wavelengths from 0.5 to 2.4\,$\mu$m. 
They allowed Mouhcine \& Lan\c{c}on (2002) and Maraston (2005)
to produce extended synthetic single stellar population spectra (SSP spectra) 
with a near-IR resolution $R=\lambda/\delta \lambda
\sim 1000$. That work focused on intermediate age populations, 
in which O-rich and C-rich TP-AGB stars are the predominant sources of 
near-IR light. Only a handful of red supergiant spectra were available at the
time. The synthetic spectra presented here represent an important step towards 
a more robust extension of previous work to ages between 
5\,Myr and a few 100\,Myr.

\subsection{New input spectra}
\label{library.sec}

The models in this paper use the empirical spectra of luminous red stars
that were compared with theoretical spectra by Lan\c{c}on et al. (2007\,a;
hereafter LHLM07). They were acquired in part with CASPIR on the
2.3m Telescope of the Australian National University (McGregor et al. 1994), 
in part with SpeX on NASA's InfraRed Telescope Facility (Rayner et al. 2003). 
All the spectra cover wavelengths from 0.9 to 2.4\,$\mu$m continuously, 
at $\lambda/\delta \lambda \geq 750$ (the SpeX data extend to 0.81\,$\mu$m). 
The flux calibration through this wavelength range has been achieved using
warm stars with known photometry. We refer to Lan\c{c}on \& Wood (2000) and
Vacca et al. (2003) for the calibration methods applied respectively to 
the CASPIR and the SpeX data. 

The theoretical spectra available to LHLM07 were computed with the
{\sc phoenix} code (P. Hauschildt and collaborators), for a metallicity
typical of the sun but with surface abundances of C, N and O modified according 
to expectations from stellar evolution calculations. We refer to that article for 
a more complete description. The authors also describe and discuss the
minimum-$\chi^2$ procedure used to estimate the effective temperature
(T$_{\rm eff}$), the surface gravity $g$ and the extinction A$_{\rm V}$ of
individual stars. The adopted extinction law is taken from Cardelli et al.
(1989), with $R_{\rm V}=3.1$. 

The conclusions of LHLM07 most relevant to
near-IR synthesis at R\,$\sim$\,1000 are the following:
theoretical spectra {\em are} able to reproduce those of static
red giants down to effective temperatures of about 3500\,K;
the models available to the authors {\em did not} as yet provide a satisfactory 
match of the spectra of cool and luminous red supergiants, even when the
effects of internal mixing on surface abundances had been taken into account.
Further theoretical work is being undertaken to improve 
the supergiant models; as yet, using theoretical spectra directly
for population synthesis purposes is considered premature.

In order to build a suitable library for population synthesis purposes,
the red supergiants and bright giants of the observed sample were subdivided 
into subsets of stars of class Ia, class Iab and class Ib/II. 
Each subset was sorted according to estimated
T$_{\rm eff}$.
Bins in T$_{\rm eff}$ were then constructed, containing
between 1 and 5 stars depending on the number of spectra available in
a given T$_{\rm eff}$ range, 
and depending on the individual quality of those data.
The spectra in each bin were examined individually, and compared with
spectra in neighbouring bins and in neighbouring luminosity classes.
This had the purpose of removing or reclassifying outliers, in
order to obtain sequences along which spectral features evolve
reasonably regularly. It must be kept in mind that,
since synthetic spectra cannot reproduce the empirical
spectra well, a large uncertainty affects estimated parameters
such as Teff or log($g$). Therefore, there is indeed quite some
freedom for adjustment, especially for temperatures below 3800\,K
and for stars of class Ia and Iab.

The spectra within a given bin were then dereddened and averaged.
All observed spectra cover
wavelengths between 0.97 and 2.4\,$\mu$m but only some extend to
shorter wavelengths. Our bin averages all reach 0.51\,$\mu$m:
the best fitting model was used to extend the data
in bins where no empirical short wavelengths spectra were available. 

\begin{figure}
\includegraphics[clip=,width=0.49\textwidth]{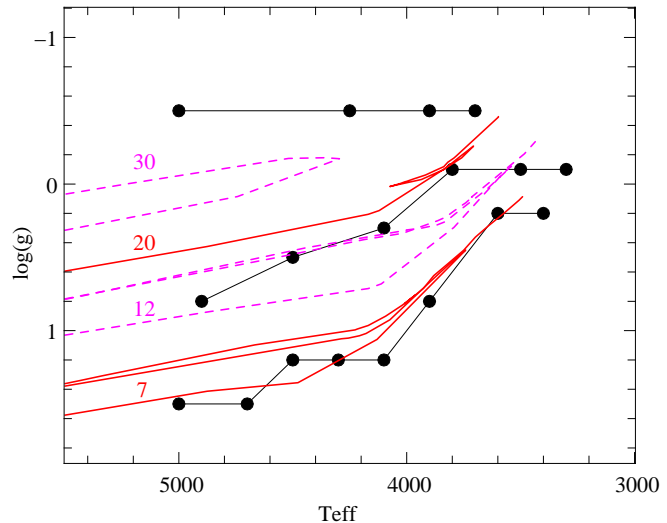}
\caption[]{Temperatures (K) and gravities (cm.s$^{-2}$)
assigned to the average 
spectra of stars of luminosity class Ia (lowest gravities), Iab (intermediate
gravities), and Ib/II (highest gravities). Overplotted are the evolutionary
tracks of Bressan et al. (1993) for the indicated stellar initial masses
(in units of M$_{\odot}$).}
\label{seq_in_HRdiag.fig}
\end{figure}

The resulting sequences have lower estimated values of log($g$)
for more luminous stars, as expected. The assigned 
values depend strongly on the assumed value of the microturbulence
parameter in the stellar models. When using microturbulent velocities of only
2 or 3\,km.s$^{-1}$ (LHLM07), 
the stars of class Ia are assigned a value of log($g$)
of $-1$, i.e. a value that is not reached by standard sets of 
evolutionary tracks. With higher microturbulent velocity
parameters (as suggested e.g. by studies of Tsuji et al. 1976, Gray et al. 2001,
Origlia et al. 1993, 1997), log($g$) rises to $-0.5$ or 0 for the 
same class Ia spectra, and the spectral fits tend to improve
(Lan\c{c}on et al. 2007\,b). 
We have taken this trend into account qualitatively
to locate the sequences in Fig.\,\ref{seq_in_HRdiag.fig}. 
At the time of this work, the grid
of large microturbulence models has too coarse a sampling in T$_{\rm eff}$
to allow us to also reassign temperatures consistently. This is 
an additional reason for uncertainty in the T$_{\rm eff}$ values along
the sequences of average spectra.

A procedure similar to the one just described has been applied
(with much smaller uncertainties on the stellar parameters) to the
empirical giant star spectra analysed by LHLW07. 
Together with the average spectra of long period
variables and carbon stars of Lan\c{c}on \& Mouhcine (2002), this
provides us with a library suitable for the construction of SSP 
spectra at both young and old ages. Details about the modelling of 
the older populations lie outside the scope of the present article.

\subsection{Population synthesis assumptions and predictions}

The synthesis of SSP spectra is performed with a version of the 
code {\sc P\'egase.2} (Fioc \& Rocca-Volmerange 1997, 1999)\footnote{{\em cf.}
{\tt http://www2.iap.fr/pegase/}} 
that has been adapted to our purposes following 
Mouhcine \& Lan\c{c}on (2002). The computation of isochrones from 
evolutionary tracks is unchanged,  but 
red supergiant phases are flagged\footnote{Red giant and asymptotic
red giant phases are also flagged but this is relevant only to intermediate age
and old populations.}.
Non-flagged points along the 
isochrones are represented with spectra of the default library
of {\sc P\'egase.2}, i.e. the semi-empirical library of Lejeune et al. (1998).
Flagged points are represented with the new average spectra of 
red supergiants described above. The bolometric
corrections needed to scale the fluxes are computed by forcing the near-IR
spectra to match the level of the Lejeune spectra in the J band. 
Other options for the bolometric corrections have not yet been explored.
For the linear interpolation between the spectra of the three red supergiant 
sequences, we choose to proceed as follows: first we interpolate along each of 
the two bracketing sequences to reach the target \Teff,
then we interpolate between the two resulting spectra according to log($g$).

The stellar evolution tracks used in this paper are taken
from Bressan et al. 1993. Our study is limited to near-solar
metallicity (tracks at Z=0.02) and we use the IMF of Salpeter (1955)
unless otherwise stated. 
Any star initially more massive than 7\,M$_{\odot}$ is flagged 
as a supergiant when it evolves off the main sequence.
As a consequence, the new collection of supergiant spectra modifies 
the predictions of {\sc P\'egase} up to an age of about 75\,Myr.  

\begin{figure}
\includegraphics[clip=,width=0.49\textwidth]{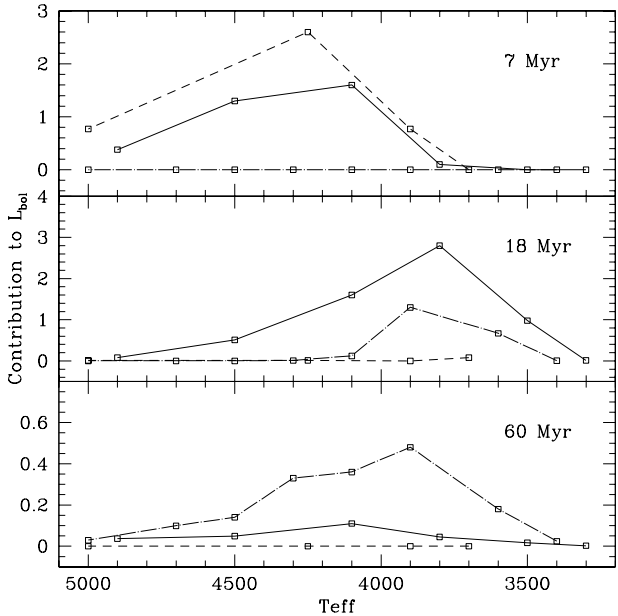}
\caption[]{Contributions of the spectra of supergiants of class
Ia (dotted), Iab (solid) and Ib/II (dashed) to the bolometric luminosity.
The contributions are average absolute values, in L$_{\odot}$, 
and are given scaled to a total initial stellar mass of 1\,M$_{\odot}$
(Salpeter IMF extending from 0.1 to 120\,M$_{\odot}$). }
\label{contribs.fig}
\end{figure}

\begin{figure}
\includegraphics[clip=,width=0.49\textwidth]{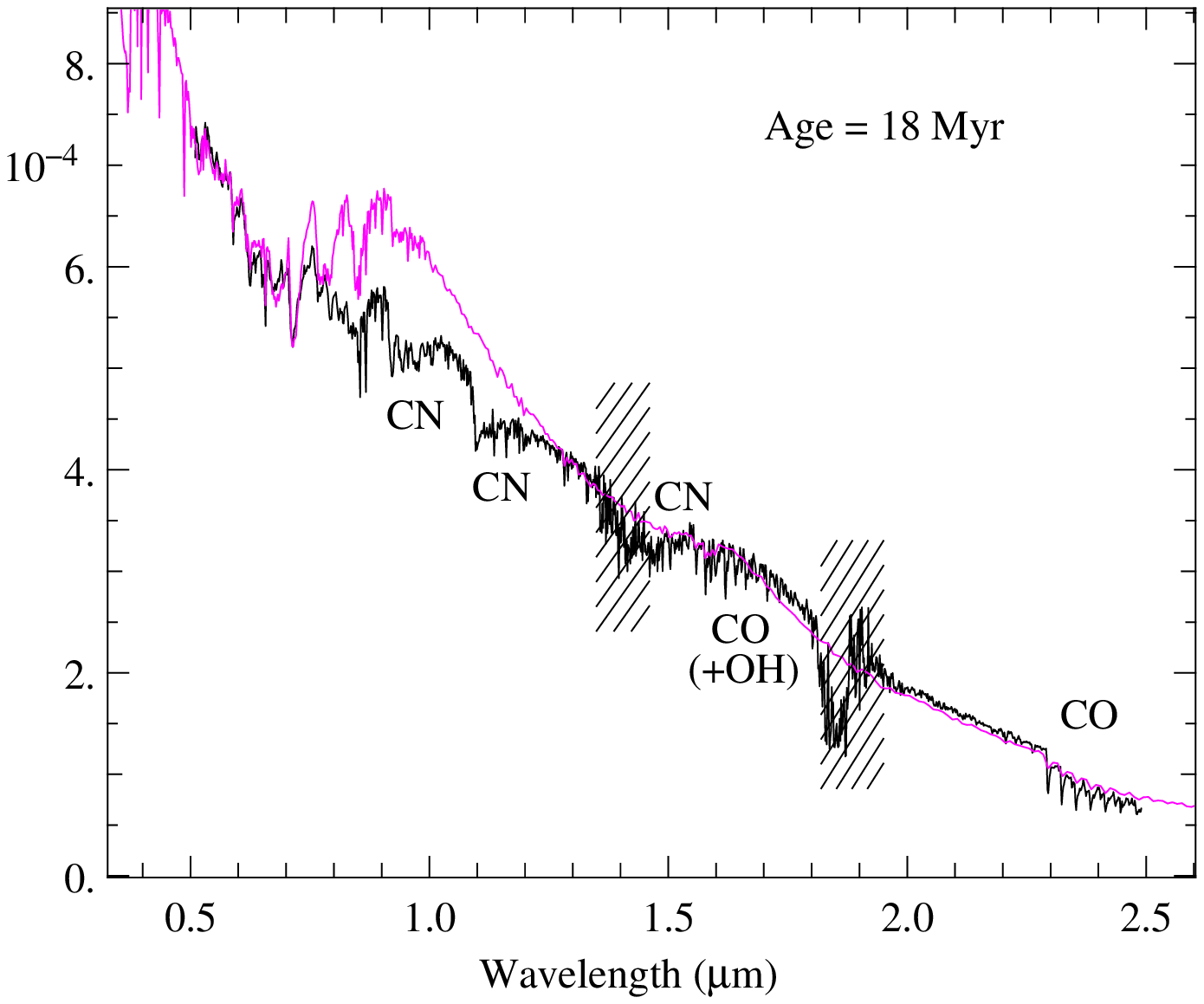}
\includegraphics[clip=,width=0.49\textwidth]{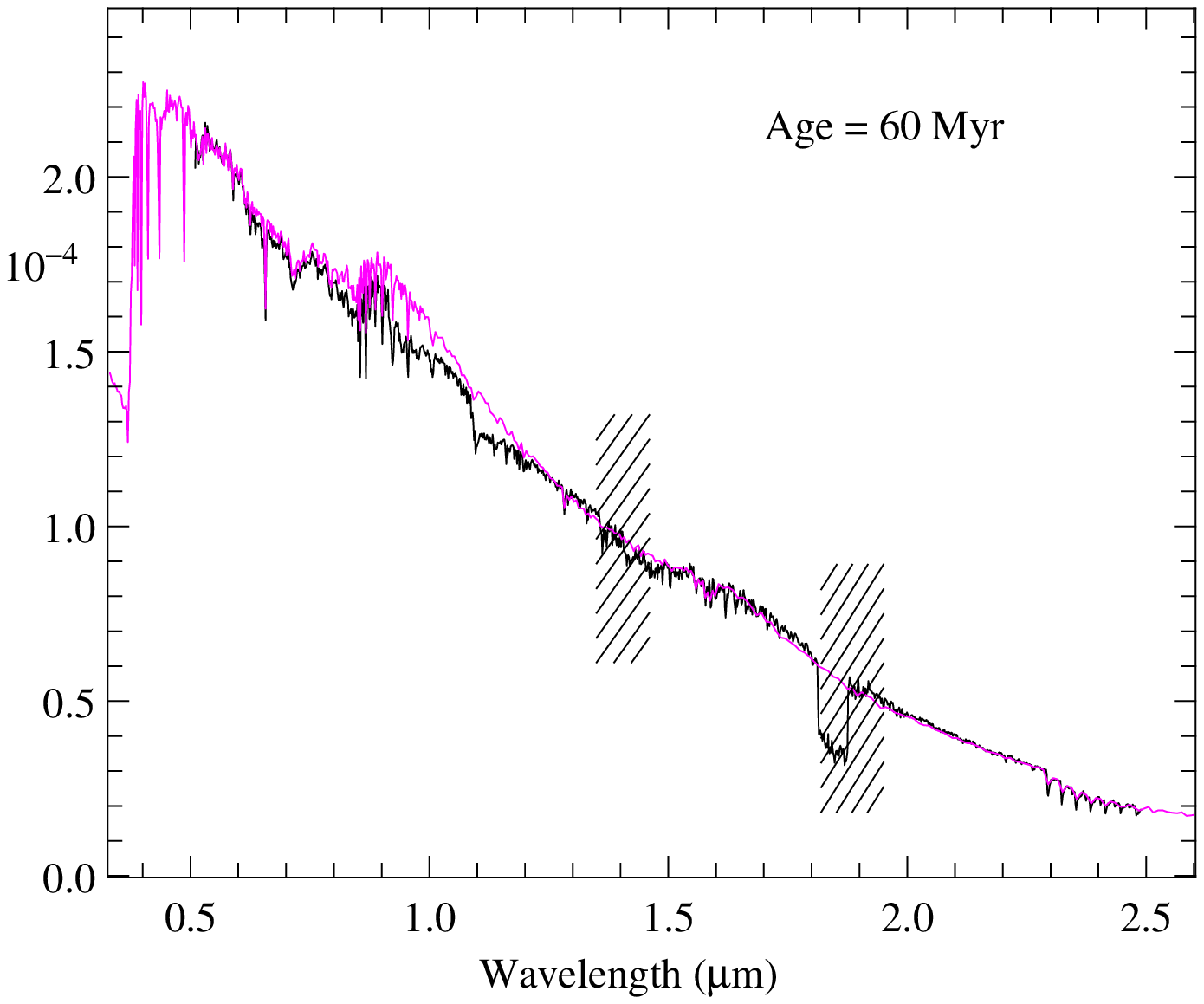}
\caption[]{Comparison of the new model spectra of single-age populations
(higher spectral resolution) with the standard predictions of {\sc P\'egase.2}
(lower spectral resolution). The spectra have been reddened for display
(top: A$_{\rm V}$=1, bottom: A$_{\rm V}$=1.5), and are presented in
arbitrary units of energy per unit wavelength. Regions of low signal-to-noise
ratio and partly missing data are shaded.}
\label{new_vs_peg.fig}
\end{figure}

Figure\,\ref{contribs.fig} indicates which of the average
red supergiant spectra contribute most strongly to the
bolometric light of single-age populations at any given ages. 
Spectra of the class Ia sequence are dominant (among red supergiants)
only at the youngest ages (7-8\,Myr). Their contribution becomes 
negligible after about 15\,Myr, but we note that this number is
particularly sensitive to the value of the gravity assigned to the stars 
observed. Stars of class Iab are dominant between ages of about 10 and
about 25\,Myr. Afterwards, the strongest red supergiant contributions
to L$_{\rm bol}$ come from stars of class Ib/II. 

The red supergiants all together are responsible for 
70--85\,\% of the 2\,$\mu$m flux and 45--75\,\% of the 1\,$\mu$m flux,
on average, at ages between 9\,Myr and 65\,Myr (see Sect. 5.2). Stars hotter than
5000\,K provide the rest of the emission. Because the library of
Lejeune et al. (1998) has a low resolution in the near-IR,  our models
will lack the high resolution spectral features of the warmer stars, and
this is a caveat in particular at short wavelengths. For the main molecular bands
in the H and K windows, implications are limited. Stars warmer than 5000\,K have very
little CO absorption at 1.6\,$\mu$m and their contributions at 2.3\,$\mu$m
remain small. 
 
As it can be seen in Fig\,\ref{new_vs_peg.fig}, the new spectra do not 
only improve the spectral resolution but they also modify the energy
distribution. With the adopted bolometric corrections (forced match with
the old models in the J band), the J, H and K magnitudes are not changed
significantly, but the flux around 1\,$\mu$m is strongly modified 
for ages between 10 and 25\,Myr. The molecular bands of CN (e.g.
1.1\,$\mu$m, 1.4\,$\mu$m) were not present at an appropriate level in
the theoretical models that entered the semi-empirical library of 
Lejeune et al. (1998).

\section{Observations and the M82 Cluster Sample}
\label{clusters.sec}

\subsection{Observations}

\begin{table}
\caption[]{Observations of star clusters in M82}
\label{obslog.tab}
\begin{center}
\begin{tabular}{cccc}
\hline \hline
Target & Date & Usable time  & P.A.  \\ 
       & (d/m/yr)(UT) & on target &        \\ \hline
L \& F & 31/3/02 & $2 \times 120$\,s & 22$^{\circ}$  \\
1a \& 1c & 3/4/02 & $2 \times 240$\,s & 125$^{\circ}$  \\
z      & 3/4/02 & $3 \times 240$\,s & 125$^{\circ}$  \\
nucleus & 3/4/02 & $4 \times 240$\,s & 22$^{\circ}$ \\ \hline
\end{tabular}
\end{center}
\end{table}

\begin{figure}
\includegraphics[clip=,width=0.49\textwidth]{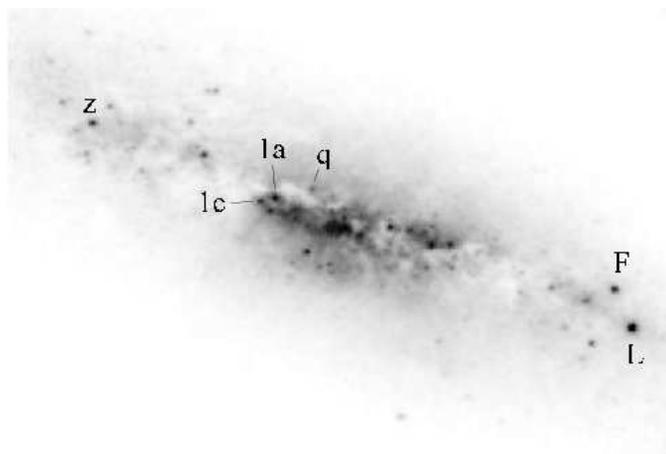}
\caption[]{Bright near-IR star clusters in the inner disc of M\,82.
Shown in inverted greyscale are two merged HST/NICMOS3 narrow H band
archive images (filter F164N; HST program 7218, P.I. Rieke). 
The clusters discussed in the present article are labelled
following the nomenclature of McCrady et al. (2003) and
McCrady \& Graham (2007). 
}
\label{M82_names.fig}
\end{figure}

Near-IR spectra of the nucleus and of five star clusters in M\,82 were acquired
with the  infrared spectrograph SpeX (Rayner et al. 2003) mounted on the NASA
Infrared Telescope Facility (IRTF), as detailed in Table~\ref{obslog.tab}.
The targets were selected to be among the brightest near-IR sources in M\,82.
They are identified in Figure \ref{M82_names.fig},
following the nomenclature of McCrady \& Graham (2007).

\begin{figure}
\includegraphics[clip=,width=0.49\textwidth]{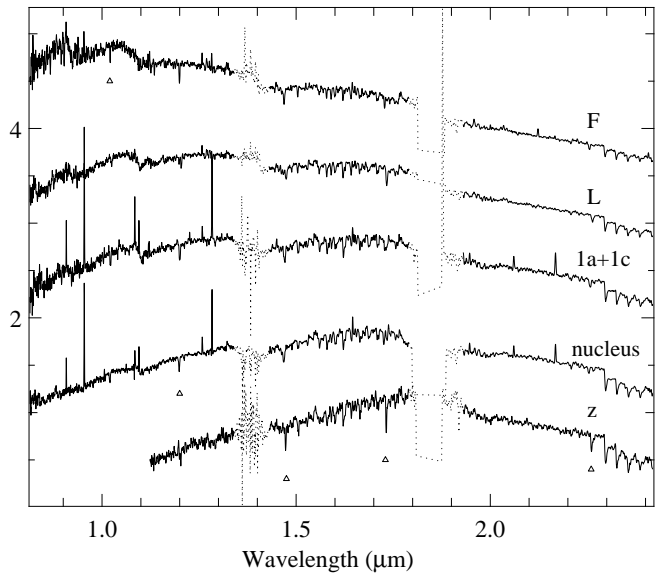}
\caption[]{IRTF/SpeX spectra of the observed star clusters in M\,82. The 
nuclear spectrum is also shown for reference. The data are normalised to
unit average flux density (energy per unit wavelength) in the H band, and
offset for clarity (the zero flux line holds for cluster $z$). They 
have been smoothed slightly for display. Triangles indicate bad pixels.
Dotted spectral segments are regions of low signal-to-noise ratio (residuals
of the telluric correction) or missing data.}
\label{allclusters.fig}
\end{figure}

The short wavelength setting of
the SpeX instrument covers wavelengths from 0.81 to 2.4\,$\mu$m
in a single observation, thus eliminating some of the uncertainties inherent
in the relative flux calibration of independent observations in the Z, J, H
and K windows. Since many of the spectral features of interest
are broad molecular bands that extend across the ``forbidden" regions
of high  telluric absorption (telluric H$_2$O around 1.15, 1.4
and 1.9\,$\mu$m), this is a significant advantage of SpeX over
many other current instruments.

Observing conditions were not photometric, and as a compromise between
spectral resolution and the need to collect photons,
we used a $0.8\arcsec \times 15\arcsec$ slit, giving a
resolving power of $R=750$. A fraction of the
red supergiant stars were observed during the same run, with the
same instrumental setting.

The targets were observed with SpeX
in a series of individual exposures of 120\,s or
240\,s. Sky exposures were obtained for each frame at 140$\arcsec$ to the
South-East of the target position, well outside the main body of the galaxy.
Several A0 type stars with known photometry were observed for the relative flux
calibration and for the correction of telluric absorption
(Vacca et al. 2003). In the end,
the observations of the A0V star HD\,92738 provided the best telluric 
correction for all the targets in M\,82. The data
reduction was based on the instrument-specific software package,
Spextools, version 3.1 (Cushing et al. 2004). 
This software offers
the possibility of rescaling the spectral segments of various orders
after they have been flux calibrated and before they are merged. But,
as expected when the calibration is successful, we found no need to
apply such an order-dependent scaling factor.

The final merged spectra are shown in Fig.\,\ref{allclusters.fig}.
They are available from the authors upon request.

\subsection{The M82 Cluster Sample}
\begin{table}
\caption{Summary of M82 Cluster Properties}
\label{props.tab}
\begin{center}
\begin{tabular}{l l l l}
\hline\hline
Cluster & Age  (optical) & Dynamical Mass & Age (IR) \\
& (Myr) & ($10^5$~M$_\odot$) & (Myr)\\
\hline\\
F & 40--80$^{1,2,3}$ & 4.7--13$^{2,3,4}$ & 15--50 \\
L  & $\approx 60^1$ & 34--46$^4$ &  10--35\\
1a & 6--7$^5$ & 7.6--9.6$^4$ & 9--25 \\
1c & & 4.4--6.0$^4$ & 9--25 \\
$z$ & & 4.7--9.45$^6$ & 10--30 \\
\hline
\end{tabular}
\end{center}

References:
1. Gallagher \& Smith (1999); 
2. Smith \& Gallagher (2001); 
3. Bastian et al. (2007);
4. McCrady et al. (2007); 
5. Smith et al. (2006);
6. see text.
\end{table}

We now describe previous optical studies of our cluster sample; a summary of their derived ages and masses is given in Table~\ref{props.tab}.

Clusters \object{M82-F} and \object{M82-L} are
located in the western disc outside the main star forming region. They
have been studied previously using optical spectra that extended to
about 9000\,\AA, and high resolution
images (Gallagher \& Smith 1999; Smith \& Gallagher 2001; McCrady et al. 2005,
Smith et al. 2006, Bastian et al. 2007).
Cluster F is intrinsically fainter than L but is seen through
a lower optical depth of dust:
it appears brighter than L in the V band but fainter in K.
Based on optical spectra the age favoured for cluster F lies between
40 and 80\,Myr (Gallagher \& Smith 1999, Smith \& Gallagher 2001, Bastian et al. 2007).
Dynamical mass estimates range from 4.7\,10$^5$ to 13\,10$^5$\,M$_{\odot}$
(Smith \& Gallagher 2001; Bastian et al. 2007; McCrady et al. 2007).
The low mass-to-light ratio of cluster F has been used to infer a
top-heavy IMF (Smith \& Gallagher 2001).

The optical spectra of cluster L indicate a similar age
(around 60\,Myr). McCrady et al. (2007) find a dynamical mass
as high as 34--46\,10$^5$\,M$_{\odot}$ for this cluster.

Clusters \object{M82-1a} (labeled A1 in Smith et al. 2006) 
and \object{M82-1c} are located less than
1$''$ apart at the edge of the main star forming region
in the inner eastern disc (region A of O'Connell \& Mangano, 1978),
in an area surrounded by very thick lanes of dust.
The optical spectroscopic age of cluster 1a is  6--7\,Myr,
i.e. significantly younger than that of L and F. No spectroscopic
age is available for cluster 1c to our knowledge.
Photometric masses for cluster 1a range from 6.4\,10$^5$ to
18\,10$^5$\,M$_{\odot}$ depending on the adopted stellar
initial mass function (Smith et al. 2006). Dynamical masses
have values of 7.6-9.6\,10$^5$\,M$_{\odot}$ for cluster 1a and
4.4-6.0\,10$^5$\,M$_{\odot}$ for cluster 1c (McCrady et al. 2007).

Cluster \object{M82-z} is located in the eastern disc region known as
region B (O'Connell \& Mangano 1978).
It is not mentioned in the optically selected cluster list for this
region by de Grijs et al. (2001),
and is also absent in the $U$-band selected cluster sample
studied by Smith et al. (2007). Indeed, cluster
$z$ is barely detected at F555W in the HST/ACS Wide Field Channel mosaic of
M82 (Mutchler et al. 2007).

The high near-IR brightness of cluster $z$ suggests that the
cluster may be in the red supergiant-dominated phase of evolution, i.e.
that it might be one of the youngest clusters in region B. Smith et al. (2007)
find that the peak epoch of cluster formation in this region was 
$\sim 150$ Myr ago, with clusters forming at a lower rate until 12--20 Myr ago.
To estimate the dynamical mass of cluster $z$, we combined the
velocity dispersion measurement of McCrady et al. (2007) with
a new radius measurement on the F814W image from the 
ACS mosaic of M82 (Mutchler et al. 2007;
measurement by I. Konstantopoulos, private communication).
We found a mass range of 4.7--9.5\,10$^5$\,M$_{\odot}$.

\section{Near-IR analysis}
\label{analysis.sec}

\subsection{Fitting procedure}
The relative quality of the model fits to the near-IR spectrum of
a given cluster is measured with a reduced $\chi^2$. 
$$ \chi^2 = 
\frac{1}{N}\ \sum_i \ W_i\,\frac{(S_i - \alpha M_i)^2}{\sigma_i^2} $$
where $i$ runs through the spectral pixels, $S$ is the empirical 
spectrum, $M$ the model under study, $\alpha$ the factor that minimises
$\chi^2$ for this model, and $\sigma$ the estimated noise.
$W$ is set to either 0 or 1. $W=0$ is used to mask
regions where strong telluric absorption leaves strong residuals
after correction. $W$ also allows us to perform fits on selected,
restricted wavelength ranges. $N$ is the number of pixels with $W=1$.
The value of $\chi^2$ is useful only to compare the quality of 
various model fits to a given spectrum. Note that it does not fulfill 
the mathematical requirements 
(such as statistical independence of the data points) that would 
allow us to interpret its numerical values in terms of likelihoods.

The models considered are SSPs with ages between 1 and about 100 Myr, reddened
using the family of extinction laws of Cardelli et al. (1989). 
The shape of these laws at near-IR wavelengths ($>$0.9\,$\mu$m) is
independent of the adopted value of R$_{\rm V}$=A$_{\rm V}$/E(B-V). 
Therefore, the estimated near-IR age is independent of R$_{\rm V}$. 
Constraints on R$_{\rm V}$ are obtained when optical data are added
to the near-IR spectrum. The estimated intrinsic (dereddened) luminosity of 
the cluster will be affected by the assumptions on the extinction law much
more than the cluster ages.

\subsection{Cluster L }

\begin{figure}
\includegraphics[clip=,width=0.49\textwidth]{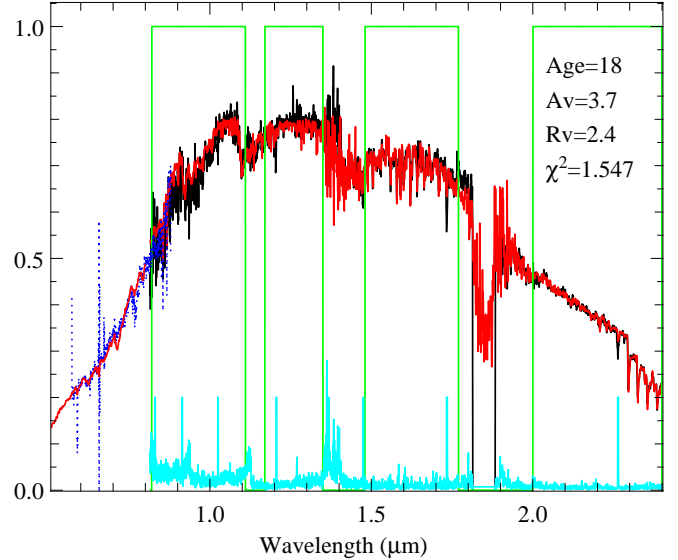}
\caption[]{Best fit to the near-IR spectrum of cluster M82-L 
(arbitrary units of energy per unit wavelength). The model (grey;
red in the electronic version)
follows the SpeX data (black) very closely over the whole range of that
data (0.81-2.4\,$\mu$m). It also matches the optical observations 
of Smith \& Gallagher (2001; dotted) if one adopts R$_{\rm V}\simeq 2.4$ 
for the extinction law. 
Wavelengths outside the rectangular boxes have zero weight
in the $\chi^2$-minimisation procedure. The noise spectrum
used to weight the $\chi^2$ is shown in light grey (cyan in the electronic
version).}
\label{L_fit1_allIR.fig}
\end{figure}

\begin{figure}
\includegraphics[clip=,width=0.49\textwidth,height=0.4\textwidth]{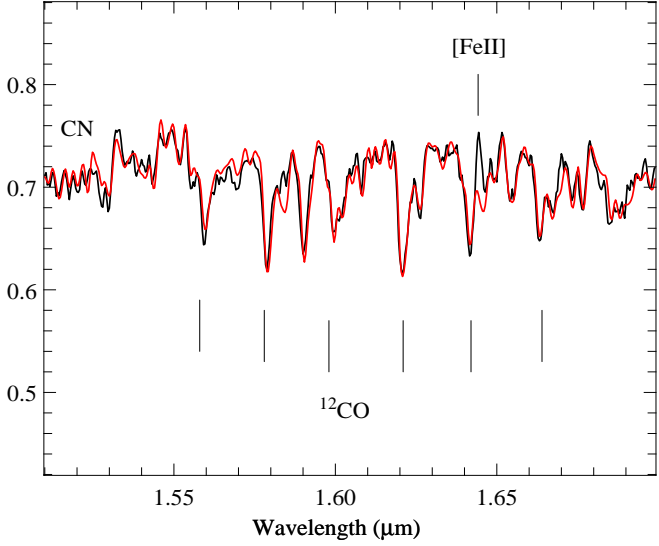}
\caption[]{Zoom on the H band of the fit of 
Fig.\,\ref{L_fit1_allIR.fig} (no additional renormalisation
was required). The fit allows for a good measurement of the [FeII]
emission line.}
\label{L_fit1_allIR_H.fig}
\end{figure}

\begin{figure}
\includegraphics[clip=,width=0.49\textwidth,height=0.4\textwidth]{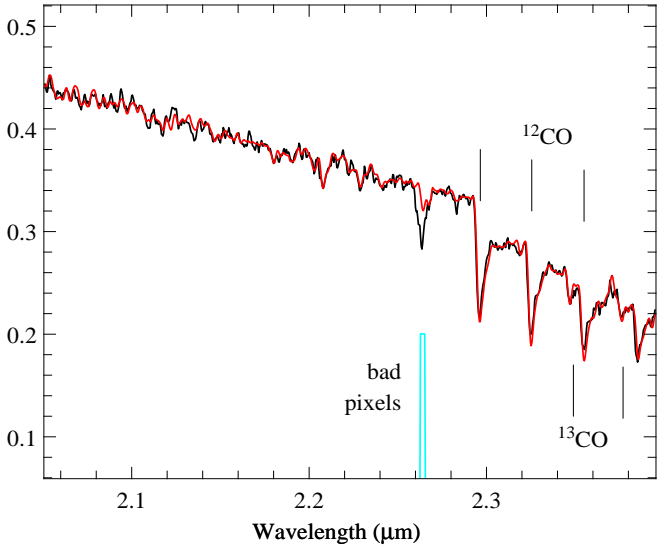}
\caption[]{Zoom on the K band of the fit of 
Fig.\,\ref{L_fit1_allIR.fig} (no additional renormalisation
was required).} 
\label{L_fit1_allIR_K.fig}
\end{figure}

Figure \ref{L_fit1_allIR.fig} shows the best fit 
to the near-IR spectrum of cluster L, 
together with optical extensions of the models and the data. The
best-fit age is 18 Myr.  Figures \ref{L_fit1_allIR_H.fig} 
and \ref{L_fit1_allIR_K.fig} show zooms on subsets of 
the spectrum: the fit quality is excellent. In particular, this fit
would provide an excellent background subtraction for the measurement
of nebular emission lines on the line of sight to the cluster. 
The extinction law of Cardelli et al. (1989) leads to 
a very satisfactory representation of the near-IR energy distribution of 
the cluster, and with R$_{\rm V}$=2.4--2.7 the fit extends
down to 6000\,\AA. 

\begin{figure}
\includegraphics[clip=,width=0.49\textwidth]{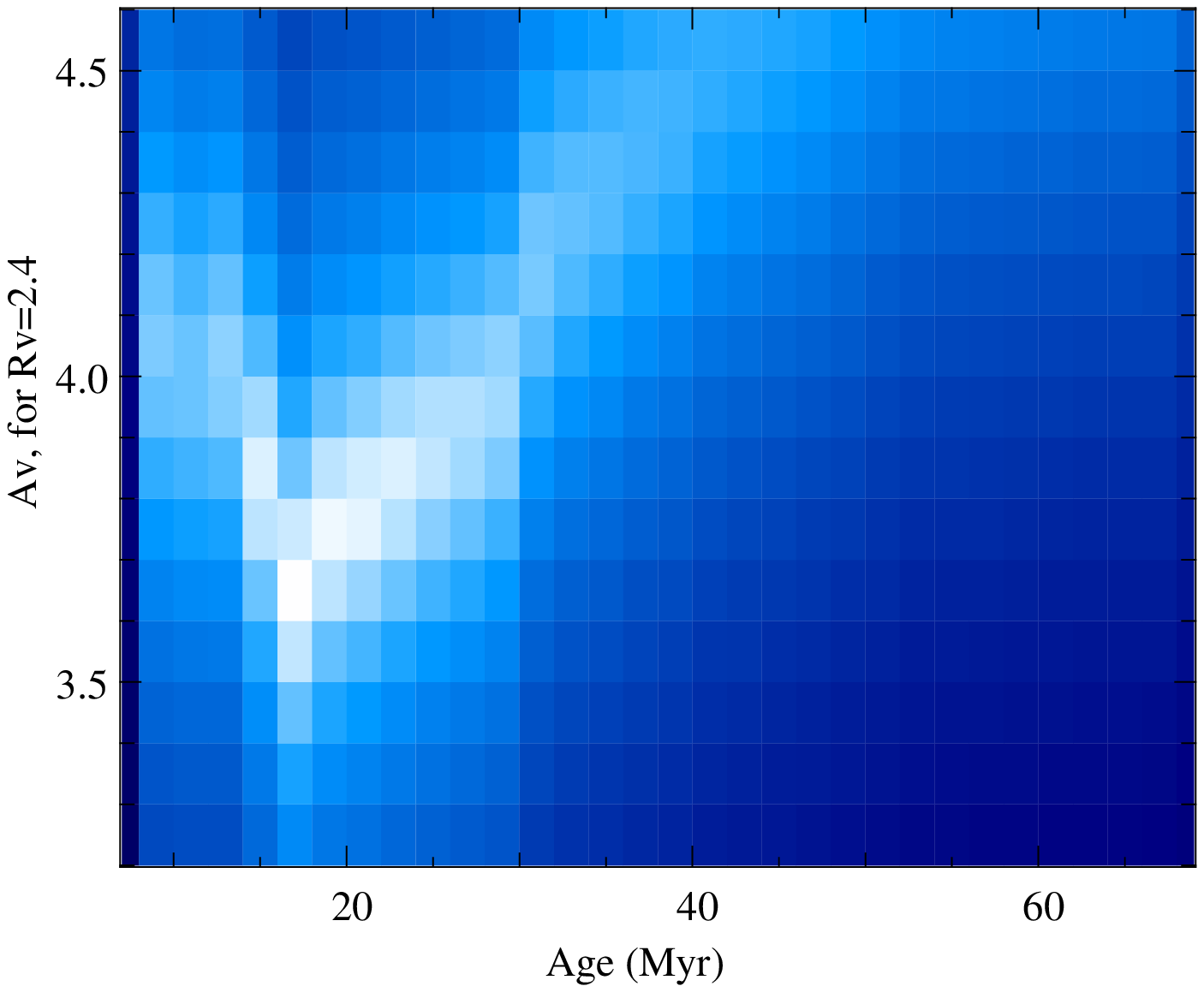}
\includegraphics[clip=,width=0.49\textwidth]{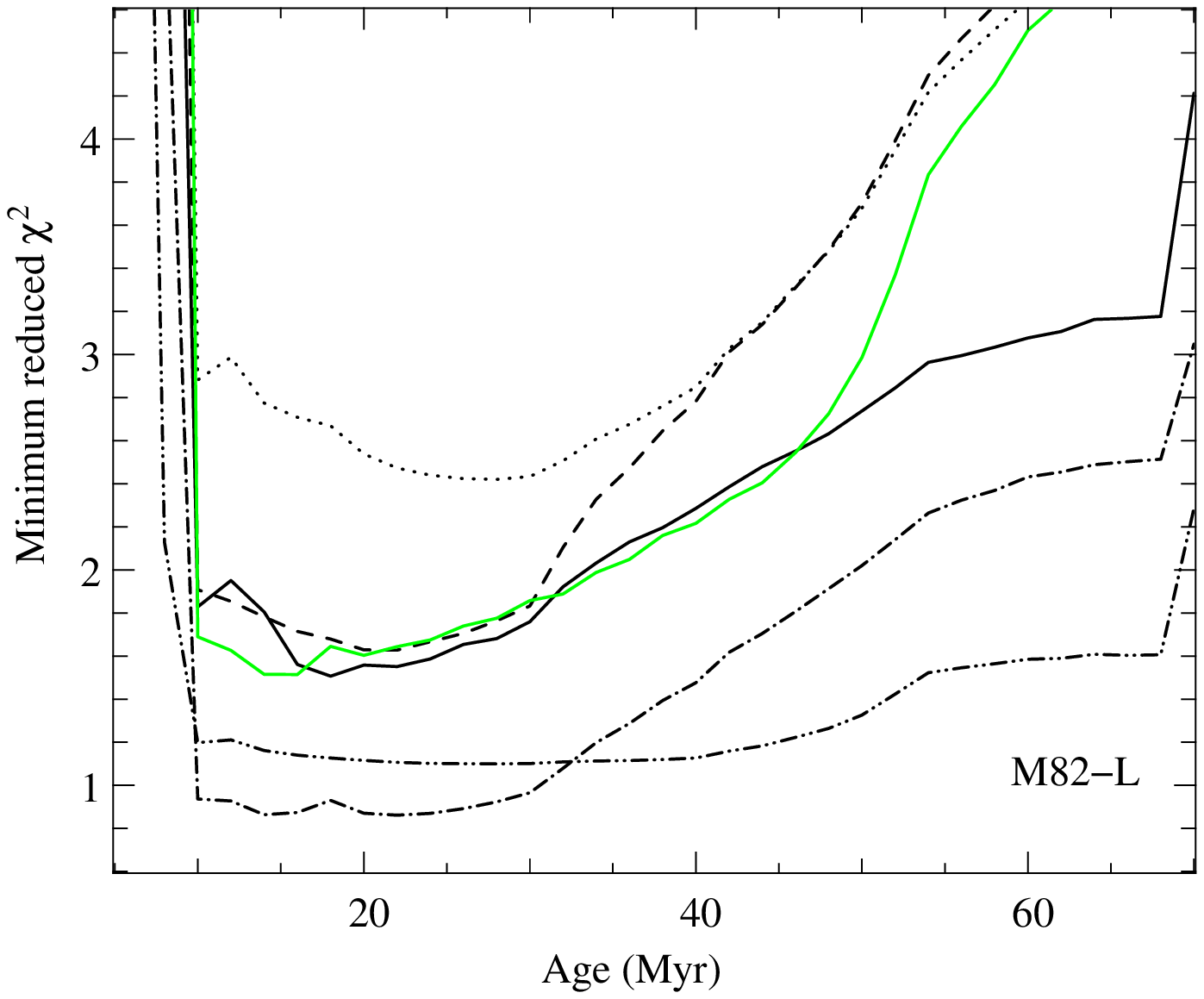}
\caption[]{Minimum $\chi^2$ as a function of age for cluster L.  
{\bf Solid:} fit performed using the full range of wavelengths, 
with weights as indicated in Fig.\,\ref{L_fit1_allIR.fig}. 
The upper graph shows the $\chi^2$ valley in the Age-Extinction plane
for this case.
{\bf Dashed:} wavelengths restricted to
the H+K windows. {\bf Dotted and dot-dashed:} wavelengths restricted 
to only the H window and only the K window. 
{\bf Dot-dot-dashed:} wavelengths below 1.35\,$\mu$m only.
{\bf Thick solid, green (or grey):} fit to the full
wavelength range but with a modified extinction law.
}
\label{L_chi2map_allIR.fig}
\end{figure}

\begin{figure}
\includegraphics[clip=,width=0.49\textwidth]{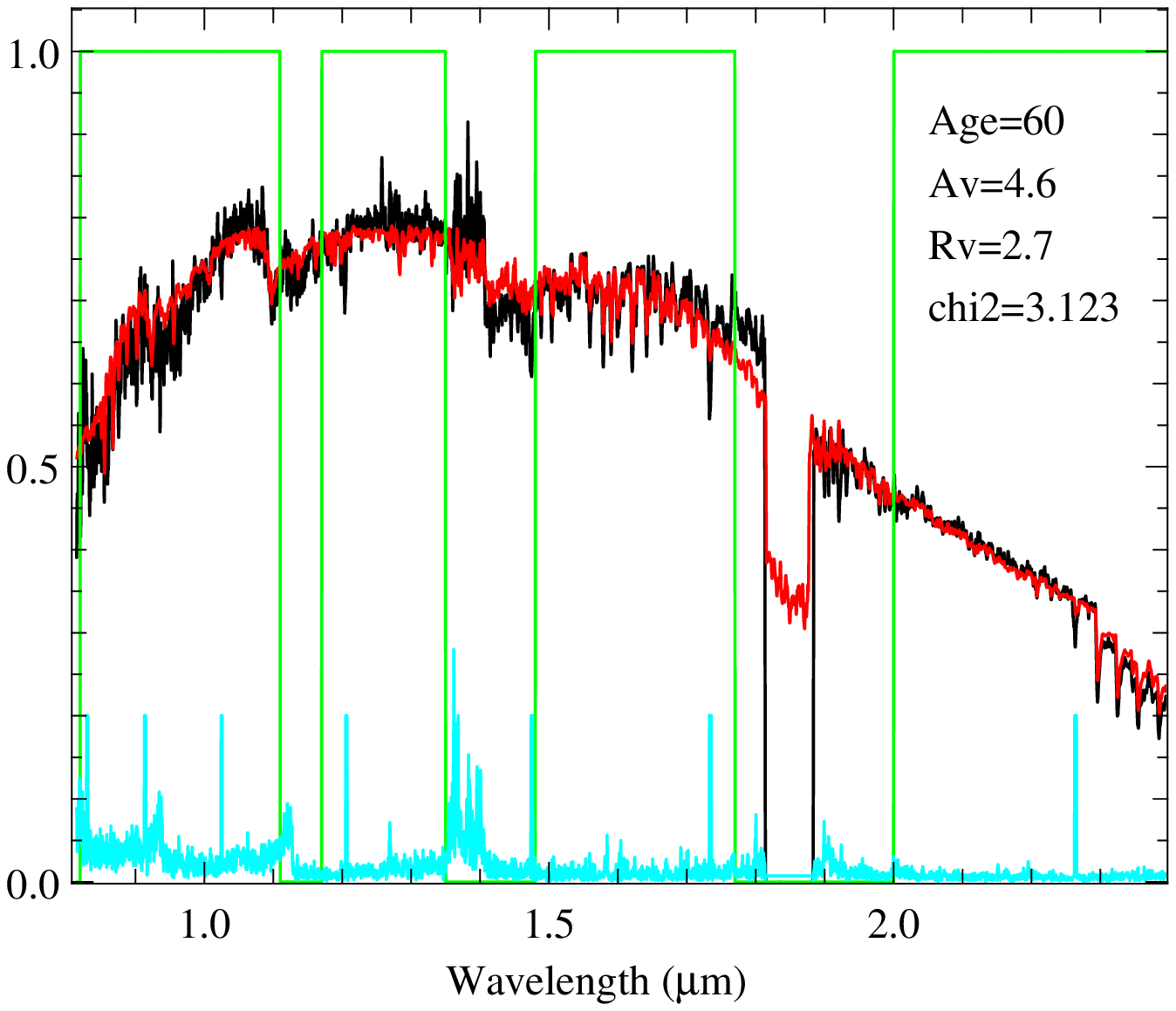}
\includegraphics[clip=,width=0.49\textwidth]{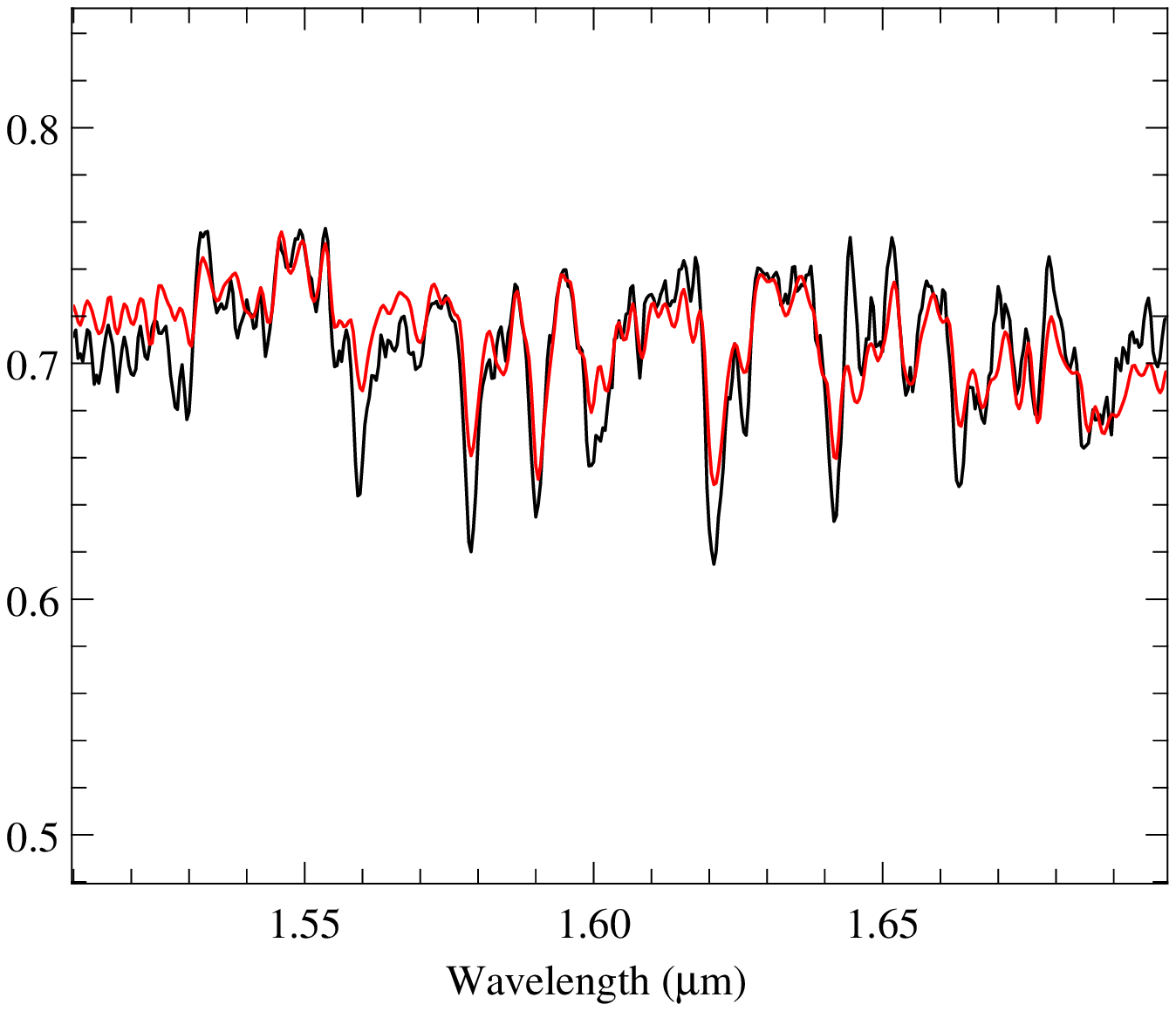}
\caption[]{Cluster L: Best fit to the near-IR spectrum and the global
energy distribution when the age is {\em assumed} equal to 60\,Myr.}
\label{L_fit60Myr.fig}
\end{figure}

The $\chi^2$-map of Fig.\,\ref{L_chi2map_allIR.fig} 
shows a valley that illustrates the
degeneracy between age and extinction present in the colours of SSPs.
The depth of the valley (solid curve in the lower graph of that figure)
measures the quality of the fit to the spectral absorption features, together
with residual effects in the reddening-corrected energy distribution. 
Separate fits to subset of the data provide a handle on uncertainties,
and have the additional advantage of being less sensitive to the actual
shape of the adopted extinction law. The figure shows that
the age ranges found in these tests overlap nicely, globally
favouring ages between 10 and 35\,Myr. Note that the
lower signal-to-noise ratio below 1.35\,$\mu$m leads to a very shallow
$\chi^2$ minimum in the corresponding curve.
One final test was performed using the full wavelength range available,
but allowing the extinction law to be modified by multiplication with
a second order polynomial. Acceptable changes in the energy
distribution were  found to be limited to less than $\sim 5\,\%$, and
the resulting age range was not significantly affected.
The value of R$_{\rm V}$ providing the best match to the global energy
distribution correlates with age: R$_{\rm V}\simeq 3.1$ is found for ages
near 10\,Myr, R$_{\rm V}<3.1$ for older ages.

\smallskip

For the sake of completeness, Figure\,\ref{L_fit60Myr.fig} shows the 
best fit obtained assuming an age of 60\,Myr, i.e. the age suggested 
by recent studies of the spectrum at wavelengths shorter than 1\,$\mu$m: 
although the energy distribution can be 
matched decently (with Rv=2.7), the fits to the molecular bands are poor. 
Within the framework of the set of SSP models described above
(i.e. the new library of red supergiant spectra, the current estimate of 
the parameters of the library stars and the adopted set of 
stellar evolution tracks), an age of 60\,Myr is excluded.
We refer to Sect.\,\ref{discussion.sec} for
a further discussion of model-dependance.

\subsection{Cluster F }

\begin{figure}
\includegraphics[clip=,width=0.49\textwidth]{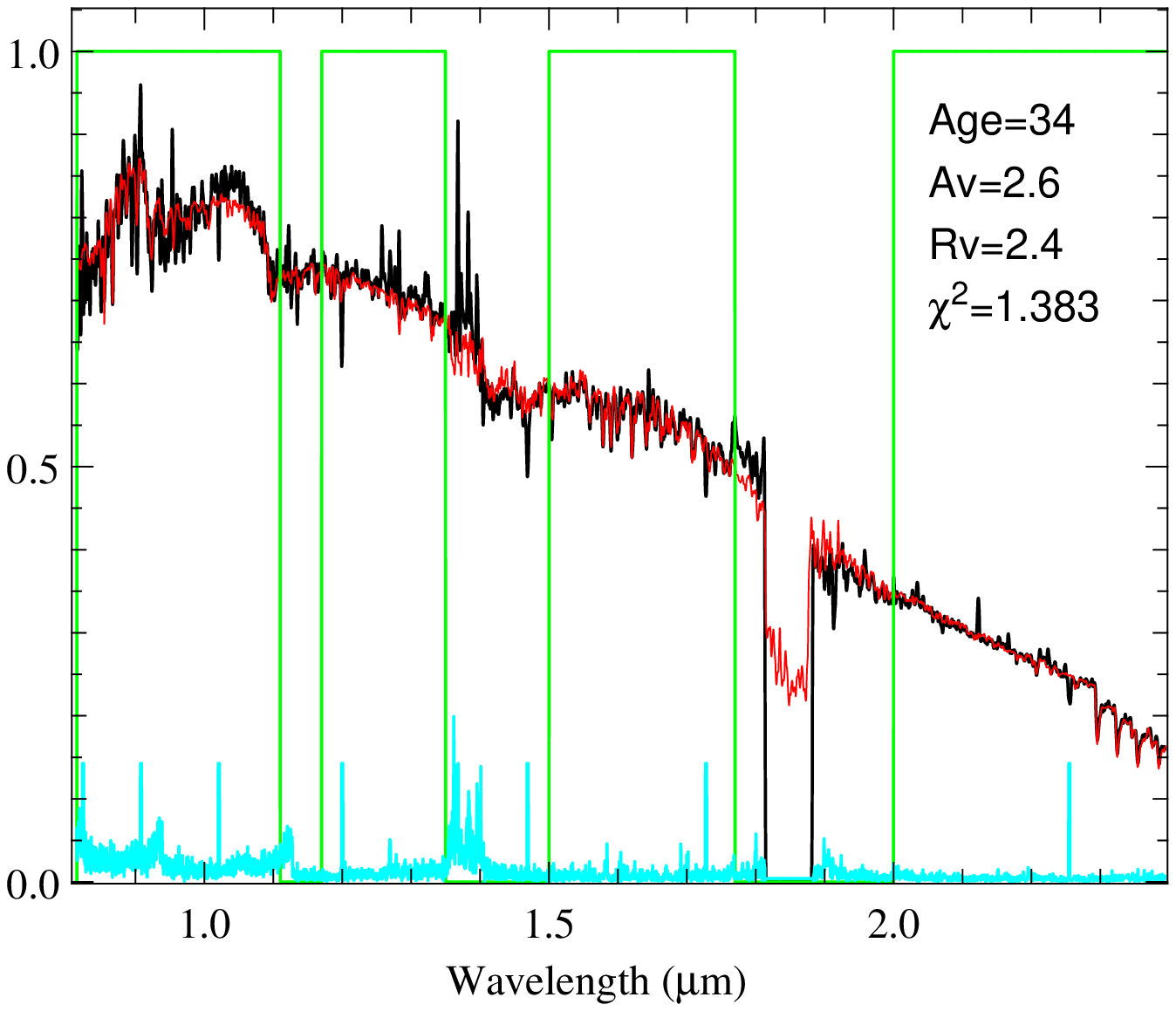}
\includegraphics[clip=,width=0.49\textwidth]{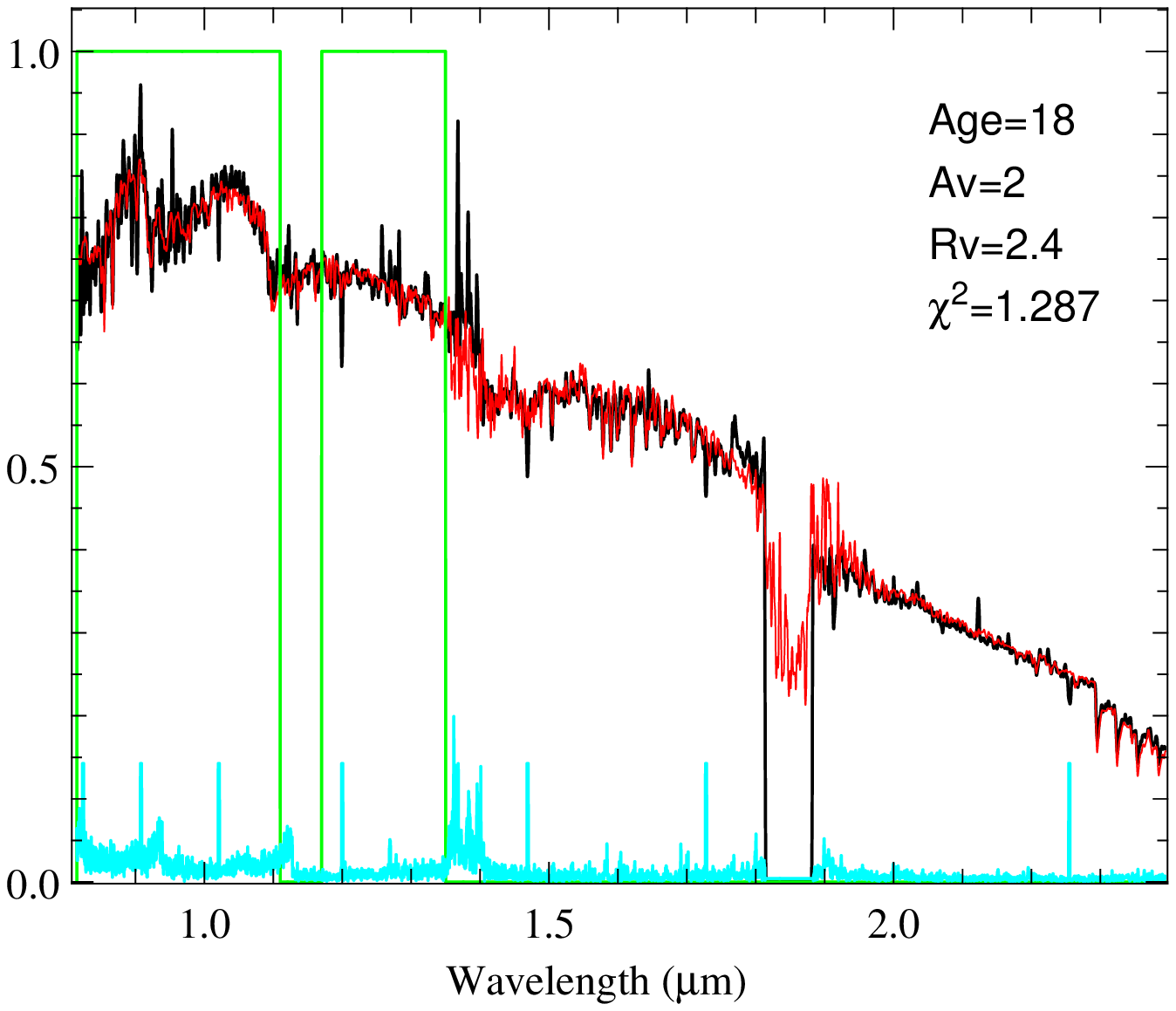}
\caption[]{Best fit to the near-IR spectrum of cluster M82-F. For the 
bottom graph, only wavelengths below 1.35\,$\mu$m have been used 
to constrain the fit. The data and the model
have been smoothed with a gaussian kernel for display (FWHM=13\,\AA),
and the plotted noise spectrum takes this smoothing into account.
}
\label{F_fit1_allIR_noV.fig}
\end{figure}

\begin{figure}
\includegraphics[clip=,width=0.49\textwidth]{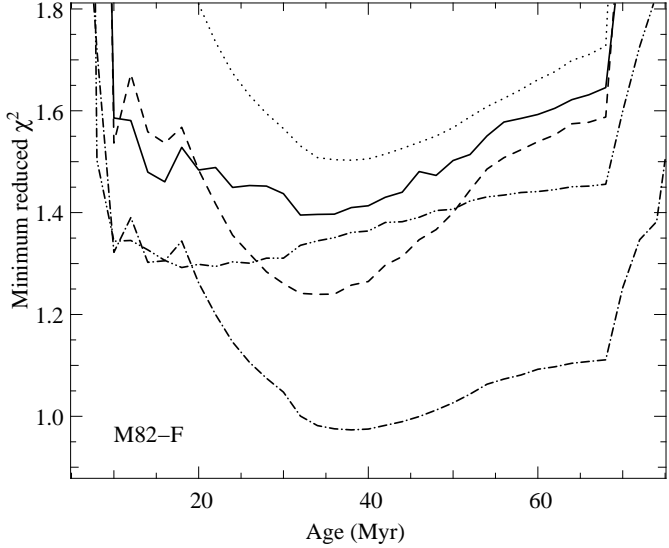}
\caption[]{$\chi^2$ versus model age for cluster F. Compare
with Fig.\,\ref{L_chi2map_allIR.fig}.}
\label{F_chi2map_allIR.fig}
\end{figure}

For cluster F, the best fit to the near-IR spectrum as a whole is
obtained for ages around 35\,Myr. The best model provides an excellent 
representation of the spectral features longwards of 1.06\,$\mu$m, 
and a good representation
of the CN bands at shorter wavelengths (Fig.\,\ref{F_fit1_allIR_noV.fig}). 
This is confirmed by fits to individual near-IR wavelength windows
(Fig.\,\ref{F_chi2map_allIR.fig}). 
The H and K band spectra alone indicate ages of 30-40\,Myr. 
A restriction to wavelengths below 1.35\,$\mu$m favours slightly
younger ages, at which SSP models display deeper CN features,
and a correspondingly lower extinction (we recall that R$_V$
affects the value of A$_V$ but not the quality of the fit 
at wavelengths $>0.9\,\mu$m). 
The bottom panel of Fig.\,\ref{F_fit1_allIR_noV.fig} illustrates
how well this younger model still reproduces the global energy
distribution of cluster F. The age-extinction degeneracy is
almost perfect: near-IR broad band colours alone could not
have provided significant constraints on either the age or the amount of 
reddening within the age range of interest.  
 
\begin{figure}
\includegraphics[clip=,width=0.49\textwidth]{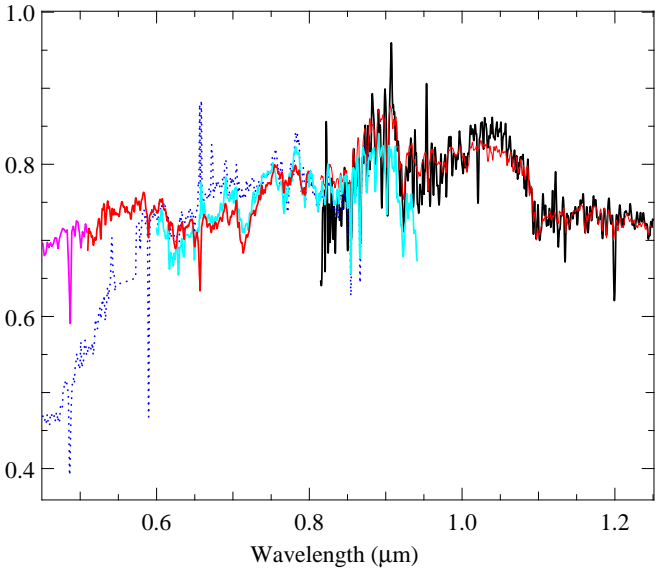}
\caption[]{Cluster F. Same as top panel of Fig.\,\ref{F_fit1_allIR_noV.fig}, 
but with available optical spectra superimposed (Age=34\,Myr,
R$_{\rm V}$=3.3 for best match to the STIS data).
The solid cyan (light grey) extension to optical wavelengths is the HST/STIS
spectrum of Smith et al. (2006), and the dotted curve is the WHT/ISIS
spectrum of Smith \& Gallagher (2001).
}
\label{F_fit1_allIR_zoomStis.fig}
\end{figure}

Cluster F has fainter magnitudes than cluster L in the near-IR and
our SpeX data, taken simultaneously with cluster L through the 
same slit, have a lower signal-to-noise ratio. As a consequence, the 
$\chi^2$-curves for cluster F have lower contrast and a broader
minimum than those for cluster L. Based on the current data and
SSP models, we derive a conservative age range of 15 to 50\,Myr
for cluster F. The CO bands around 1.6\,$\mu$m and CN bands of 
the 60\,Myr model are uncomfortably weak compared to the observations.
The near-IR age range overlaps with the younger of the optical 
ages from previous studies (Table~\ref{props.tab}).
\smallskip

In Fig.\,\ref{F_fit1_allIR_zoomStis.fig}, 
optical spectra obtained with WHT/ISIS 
(Smith \& Gallagher 2001) and HST/STIS (Smith et al. 2006)
are shown. The model displayed corresponds to the best fit to the
SpeX spectrum (34\,Myr). R$_V=3.3$ is chosen for a best match
to the slope of the STIS spectrum. The STIS spectrum can be 
matched reasonably well together with the SpeX spectrum.
In agreement with results from the SpeX data restricted to short wavelengths, 
a direct fit to the STIS spectrum alone favours the younger of the ages in
the range already indicated. 
If future improvements of the model inputs
lead to stronger near-IR molecular bands at older ages, a  more
satisfactory match between optical and near-IR ages might be obtained
(see Sect.\,\ref{disc.models.sec}).

No combination of A$_{\rm V}$
and R$_{\rm V}$ provides a satisfactory fit to the more extended combined 
energy distribution of the SpeX and ISIS data. 
Efforts to improve the models should be accompanied 
with new observations of cluster F.
Indeed, the direct comparison of the HST/STIS and WHT/ISIS 
spectra shows differences in the energy distribution, 
in the absorption features and in the strength of the nebular emission lines
(Fig.\,\ref{F_fit1_allIR_zoomStis.fig}). 
Considering the wealth of small scale structure around cluster F 
(McCrady et al. 2005, Bastian et al. 2007), 
it is not surprising that optical spectra taken one with HST/STIS, the
other from the ground differ.
The near-IR SpeX spectrum, taken under poor 
seeing conditions through a relatively broad slit, covers a different area
again. The ground based data could be contaminated by emission from younger
stars on a neighbouring line of sight, which would also explain the 
presence of weak emission lines. The complex
dust configuration could produce an obscuration law that differs
from the ones we have explored here.

\subsection{Luminous masses of clusters L and F}

\begin{figure}
\includegraphics[clip=,width=0.49\textwidth]{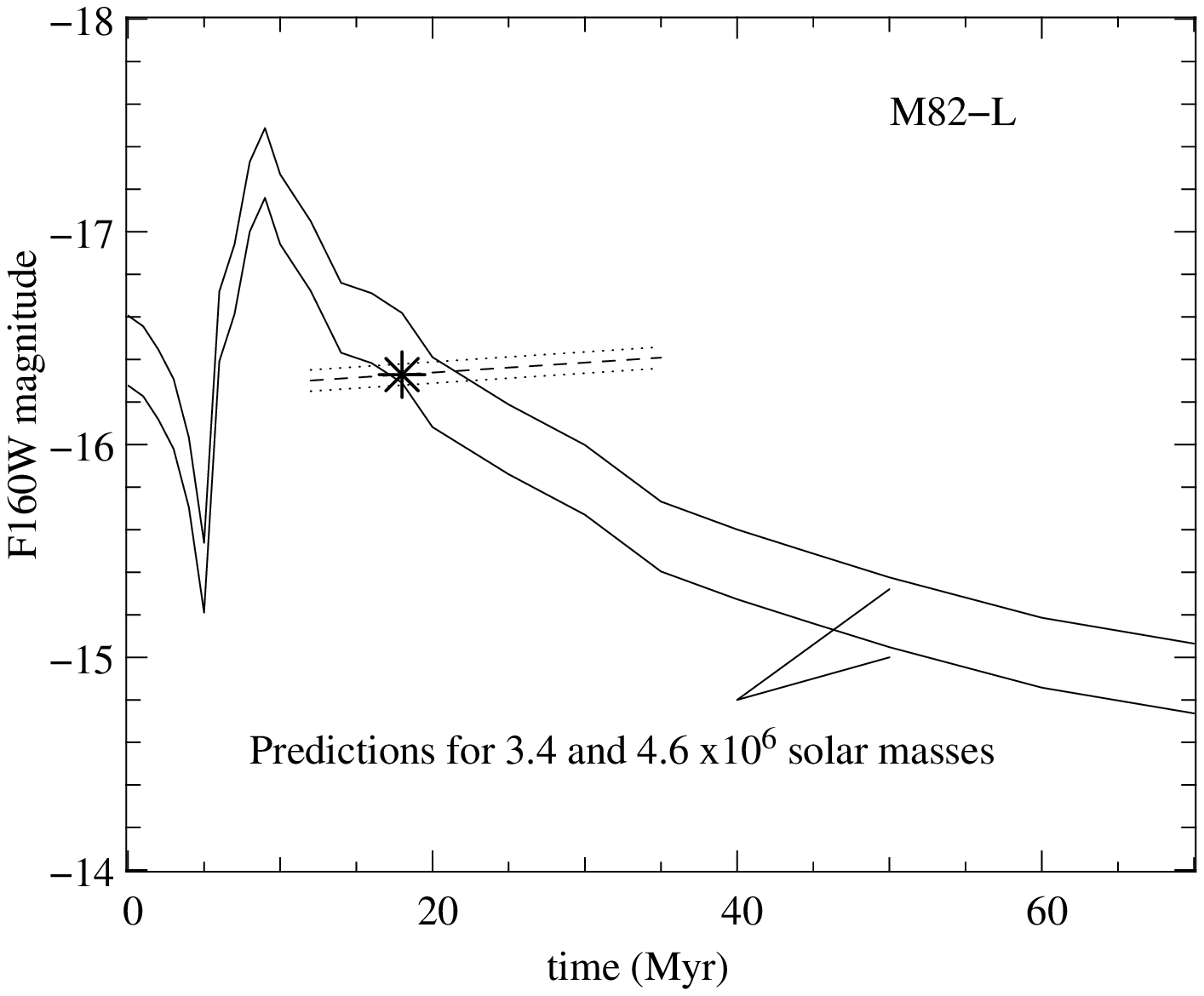}
\includegraphics[clip=,width=0.49\textwidth]{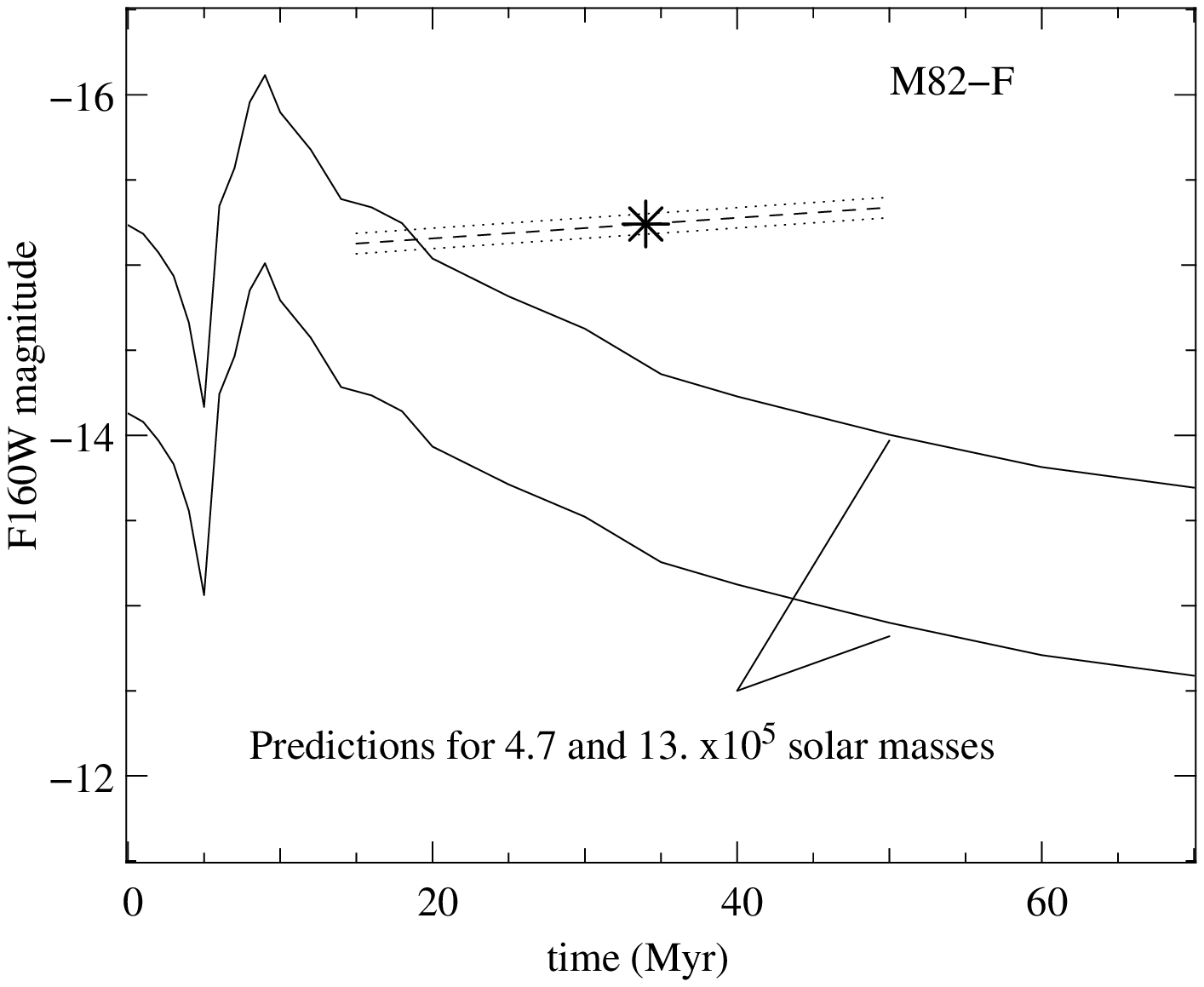}
\caption[]{Absolute dereddened magnitudes of clusters L (top) and F (bottom).
The asterisk shows the values favoured based on the near-IR analysis; 
dotted and dashed lines show the range of acceptable values. 
The solid lines are predictions, for masses that
brackett the range of dynamical masses given in Tab.\,\ref{props.tab}. A 
Salpeter IMF (0.1--120\,M$_{\odot}$) is assumed. 
}
\label{MsurL.fig}
\end{figure}

Based on the above results, we may estimate the luminous masses of clusters
L and F and compare them with dynamical mass estimates. We adopt the 
cluster magnitudes of McCrady \& Graham (2007), and assume a distance
to M\,82 of 3.6\,Mpc. We correct these magnitudes for extinction 
with the simplifying assumption that our measurement of the reddening
indeed translates into extinction as in the dust screen model of 
Cardelli et al. (1998). In Fig.\,\ref{MsurL.fig}, the measured 
F160W magnitudes (which depend on age through 
the age-dependent estimate of extinction) are compared to predictions
for single stellar populations with total stellar masses equal
to the dynamical masses of Table\,\ref{props.tab}. Our results for
cluster F are consistent with previous work: the cluster indeed 
seems overluminous. At the favoured age of 34\,Myr, it is a factor
of 2 more luminous than the prediction based on a Salpeter IMF
and on a mass of $1.3 \times 10^6$\,M$_{\odot}$. 
For cluster L, the luminous and dynamical masses agree.

\subsection{Clusters 1a and 1c }

\begin{figure}
\includegraphics[clip=,width=0.49\textwidth]{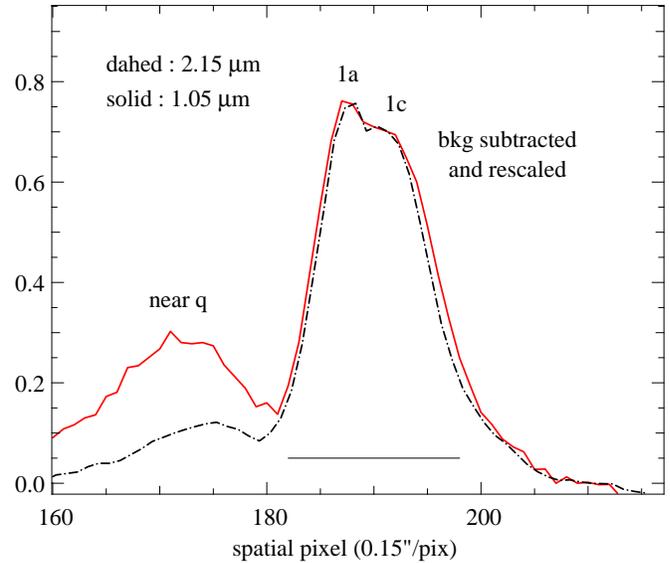}
\caption[]{Spatial profiles across clusters 1a anc 1c along the slit of
the SpeX spectrograph (after background subtraction). 
The extracted aperture is indicated. 
}
\label{T1_profiles.fig}
\end{figure}

Because of poor seeing at the time of the observations,
the spectra of clusters 1a and 1c are severely blended 
and only a combined spectrum has been extracted finally. 
It displays all the strong molecular bands typical of luminous red supergiants.
Figure\,\ref{T1_profiles.fig} shows the spatial profiles measured 
along the slit in two spectral orders, 
around 1.05\,$\mu$m and around 2.15\,$\mu$m.
The relative contributions of clusters 1a and 1c are similar at
all extracted wavelengths, suggesting either similar ages for both clusters
or a fortuitous compensation of an age difference with a difference in 
extinction. Therefore, we explore single-age population 
models before considering more complex combinations.

\begin{figure}
\includegraphics[clip=,width=0.49\textwidth]{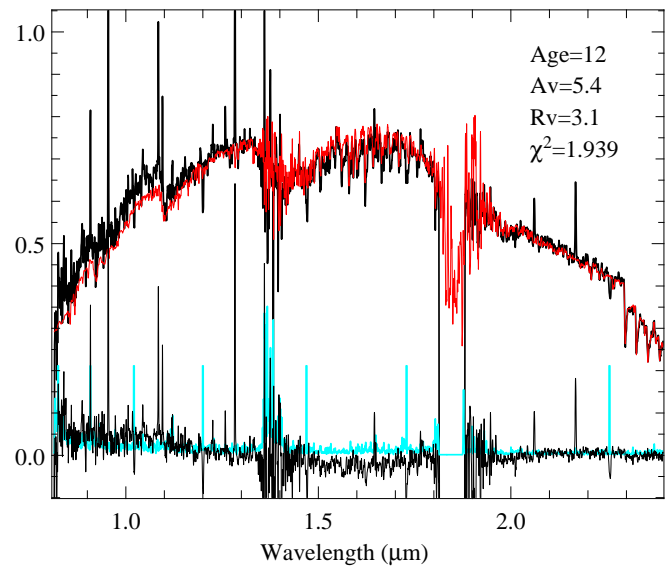}
\caption[]{Best fit to the combined near-IR spectrum of clusters 1a and 1c
with a single-age model.  
The residuals are shown in black, the noise spectrum in cyan (light grey). 
The weights
are similar to those shown for clusters F and L, except that they also 
mask emission lines.
The data and the model
have been smoothed with a gaussian kernel for display (FWHM=10\,\AA),
and the plotted noise spectrum takes this smoothing into account.
}
\label{A1_fit1_nocurve.fig}
\end{figure}

When fitting the spectrum with synthetic spectra of single-age populations
(Fig\,\ref{A1_fit1_nocurve.fig}), 
the residuals show that the global curvature of the
observed near-IR energy distributions are not matched quite as well
as in the case of cluster L or cluster F. 
The best fitting models are peaked in the H band more than the data. 
The residuals are of the order of 10\,\% only. Considering the high extinction
towards the observed sources (A$_V \sim 5$), we cannot exclude that
such a difference may be due to an inadequacy of the adopted extinction law. 
An excellent fit is obtained when the extinction curve is modified
with an {\em ad hoc} second order polynomial (which takes values 
between 1 and 1.1).

\begin{figure}
\includegraphics[clip=,width=0.49\textwidth]{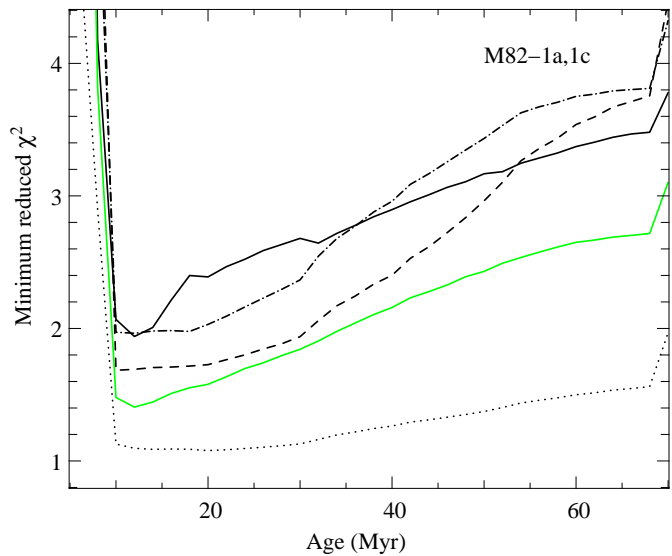}
\caption[]{$\chi^2$ constraints on the age of clusters 1a and 1c. 
{\bf Solid:} fit of the combined spectrum, 
performed using all available wavelengths, with weights similar to those
used for clusters L and F. {\bf Dashed:} wavelengths restricted to 
the H and K windows. {\bf Dotted (dot-dashed):} wavelengths restricted to the
H window (K window). {\bf Thick solid, green (or grey):} fit to all the 
available wavelengths but with a modified extinction law.
}
\label{A1_chi2map1d.fig}
\end{figure}

Figure\,\ref{A1_chi2map1d.fig} shows how the reduced $\chi^2$ varies with
model age, depending on the wavelength range used to constrain the fit.
The spectrum favours ages between 9 and 30\,Myr, i.e. the ages at 
which the red supergiant features in the model spectra are strongest.
The thick grey curve (green in the colour version of the figure) 
is obtained with the modified extinction law: the value of the best
$\chi^2$ is reduced but the behaviour with age is otherwise unchanged.

\begin{figure}
\includegraphics[clip=,width=0.49\textwidth,height=0.4\textwidth]{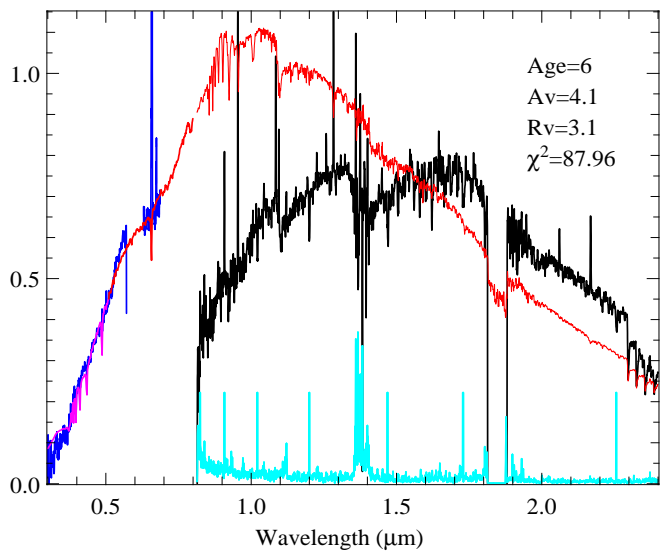}
\caption[]{Comparison of the combined SpeX spectrum of clusters 1a and 1c
(black) with a model spectrum (red) based on the age and extinction derived 
from the HST/STIS spectrum of cluster 1a by Smith et al. (2006):
age=6.4$\pm$0.5\,Myr, E(B-V)=1.35$\pm$0.15, R$_V$=3.1.
The HST/STIS spectrum of 1a is shown in blue.
Within the framework of the population synthesis models of this paper,
an age this young is not compatible with the near-IR data.
}
\label{A1_6Myr.fig}
\end{figure}

The analysis of the HST/STIS spectrum of cluster 1a 
by Smith et al. (2006) led to an age of 6 to 7\,Myr
with E(B-V)=1.35 and R$_V$=3.1 (we will refer to these estimates as
the STIS-parameters below). 
An age this young is excluded in the framework of our near-IR 
analysis (Fig.\,\ref{A1_6Myr.fig}):  at ages younger than 9\,Myr, 
the stellar evolution tracks of Bressan et al. (1993)
do not allow cool supergiants to contribute enough to 
explain the deep molecular bands we see. 

Could two-component models offer a satisfactory solution to this
discrepancy? In other words, could a model for cluster 1a based
on the STIS-parameters be combined with an older model for cluster 1c
in such a way as to match the SpeX data? 
Within our current setting, the answer is no. 
The difference in colour between the model shown in 
Fig.\,\ref{A1_6Myr.fig} and the SpeX spectrum is incompatible with 
the small wavelength dependence of the spatial profile across the 
two clusters along the SpeX slit. The profile also excludes a solution
in which an older, highly reddened cluster 1c would outweigh a 
bluer contribution from cluster 1a by large factors.

Our interpretation of the above discrepancy between optical
and near-IR ages is that it is due to an inadequacy of the
adopted set of stellar evolution tracks. Despite the impressive 
mismatch in Fig.\,\ref{A1_6Myr.fig}, the difference between derived 
optical and near-IR ages is actually small. Between ages of 5\,Myr and
10\,Myr, the evolution in the ratio between red and blue stars
is extremely rapid, and therefore also particularly model dependent. 
Fig.\,2 of Fioc \& Rocca-Volmerange (1997) shows that the predicted rapid 
transition from blue to red (V-K) colour occurs about 3\,Myr earlier with the
tracks of Meynet et al. (1994), used by Smith et al. (2006),
than with those of Bressan et al. (1993), used here. 
Tracks that include stellar rotation, or tracks
with slightly different metallicities, also modify
the time at which first red supergiants appear by a few Myr.

\subsection{Cluster z}

\begin{figure}
\includegraphics[clip=,width=0.49\textwidth]{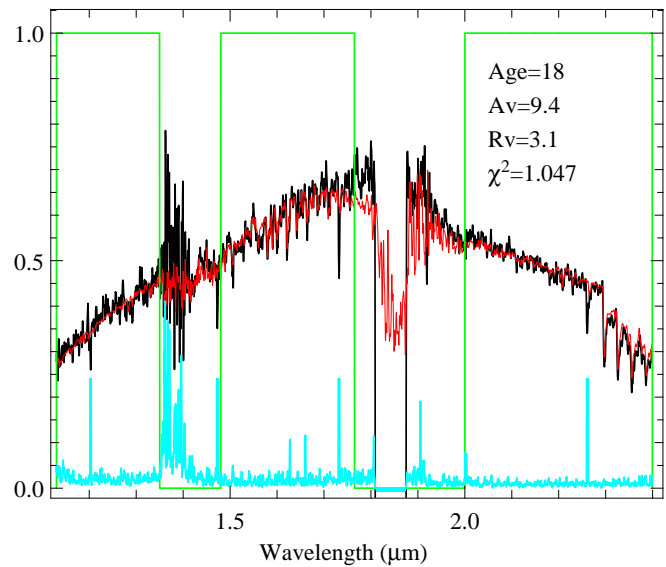}
\caption[]{Best fit to the near-IR spectrum of cluster $z$.
The data and the model
have been smoothed with a gaussian kernel for display (FWHM=12\,\AA),
and the plotted noise spectrum takes this smoothing into account.
}
\label{z_fit1.fig}  
\end{figure}

\begin{figure}
\includegraphics[clip=,width=0.49\textwidth]{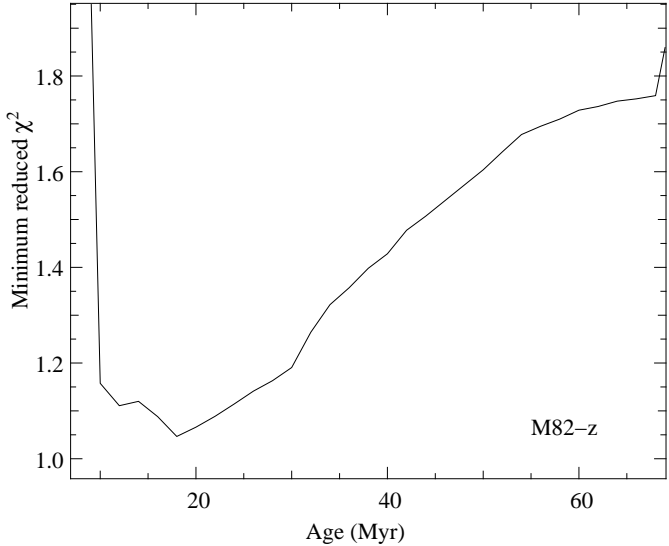}
\caption[]{$\chi^2$ constraints on the age of cluster $z$, using all
near-IR wavelengths indicated by the rectangular weight function 
in Fig.\,\ref{z_fit1.fig}. 
}
\label{z_chi2map1d.fig} 
\end{figure}

Because of the extreme extinction towards cluster $z$, 
we were able to extract its SpeX spectrum only for $\lambda > 1.13\,\mu$m.
Figures \ref{z_fit1.fig} and \ref{z_chi2map1d.fig} show the best fit
obtained and the corresponding $\chi^2$ curve. The near-IR constraints 
on age are similar to those obtained for cluster L and 1a\,: the
molecular bands are strong and this selects ages at which the 
contributions of luminous red supergiants are important. Ages between
10 and 30\,Myr are favoured. A closer look at the fits to the CO bands
in the H and K window shows that even these young models produce 
CO bands that are marginally too weak compared to the data. Ages older
than 40\,Myr are excluded in the framework of our current models.
Clearly, cluster $z$ belongs to the very youngest objects in 
region B of M\,82.

It is worth noting that the energy distribution through the near-IR range
is extremely well reproduced with the extinction law of Cardelli et al.
(1989), despite the high optical depth (A$_V \simeq 9.4$ if R$_V = 3.1$).

\section{Discussion}
\label{discussion.sec}

\subsection{Near-IR modelling}
\label{disc.models.sec}
We have shown that it is possible to obtain very good 
representations of the near-IR spectra of young massive star
clusters. 
The near-IR ages summarised in Table~\ref{props.tab},
however, do not always agree with previous optical ages. 
For the M82 clusters under study, a general trend seems to apply:
the near-IR spectra display deep molecular bands, and
as a result they point to model ages at which the contributions of
luminous red supergiants is important. 
\medskip

The input that most strongly affects the red supergiant contributions
at any given age is the adopted set of evolutionary tracks. At
ages between 5 and 100\,Myr, large differences
between the predictions for various available sets are 
obvious in colour plots, colour-magnitude diagrams and optical spectra
(Fioc \& Rocca-Volmerange 1997, Bruzual \& Charlot 2003, Levesque et al. 2005,
Gonz\'alez Delgado et al. 2005). The tracks of the Geneva group
(Schaller et al. 1992, Meynet et al. 1994,
non-rotating tracks of Meynet \& Maeder 2003)
produce more and cooler red supergiants at young ages (6--20\,Myr)
than those of the Padova group, used here. Near-IR models based on the 
Geneva tracks may well be able reconcile the optical and near-IR ages of 
cluster M82-1a. Stronger red supergiant contributions are also obtained
at a variety of ages with tracks that assume newborn stars rotate
(Meynet \& Maeder 2000, 2003; V\'azquez et al. 2007). 
Along rotating tracks, the stars tend to spend more time in redder parts
of the HR-diagram, the red supergiants evolve at higher luminosities and
lower (average) surface gravities, and they have strongly
enhanced surface abundances of nitrogen. Both optical and near-IR 
studies would be affected by these changes. 
\medskip

The second model input of importance for stellar
population synthesis is the spectral library. 
In Sect.\,\ref{library.sec}, we noted that until the stellar atmosphere
models are able to properly reproduce the extended spectra of individual
red supergiants, the values of T$_{\rm eff}$ and log($g$) assigned to
individual stars will remain uncertain. They are affected by 
metallicity, detailed surface abundances, microturbulence and
other model atmosphere ingredients.
As an illustration of the sensitivity of age dating techniques to 
these inputs we have computed SSP models in which the spectral
sequences of Fig.\,\ref{seq_in_HRdiag.fig} were shifted up (case A) 
or down (case B) in $g$ (by 0.4\,dex for class Ia, 0.2\,dex for class Iab). 
When assigned gravities are larger (case B; expected if microturbulent
velocities are large), spectra of the more luminous sequences 
are used with larger weights and over a larger range of ages than
when assigned gravities are small (case A). The strongest effect is 
a prediction of stronger CN bands with larger assigned gravities.

We confronted the resulting SSP spectra with observations of cluster F,
for which the bluer parts of the near-IR spectrum (predominantly
shaped by CN bands) 
indicated slightly younger ages than the H and K band spectra
(Fig.\,\ref{F_chi2map_allIR.fig}).
This situation suggested that models with stronger CN bands would
provide a better global fit.
As expected, case B models led to a wavelength independent 
near-IR age of 25-30\,Myr, while case A made it impossible
to match CN and CO bands simultaneously. 

This test shows that with the data quality achievable today
we {\em are} sensitive to spectral
differences between {\em luminosity subclasses},
and it will be useful to continue to include this distinction in 
future population synthesis models.

The derived ages are expected to be sensitive also to the 
adopted T$_{\rm eff}$ scale.  Assigned temperatures rise by up
to 300\,K when the microturbulence parameter is increased
to 10\,km\,s$^{-1}$, but
(based on a very preliminary exploration of a new series of
theoretical spectra) they drop by similar amounts when switching
from the solar abundances of Anders \& Grevesse, used here (see LHLM07), to
those of Asplund et al. (2005). 

In view of the above, it is clear that it will be necessary to
reassess the cluster data with a wider range of models. The large
systematic errors due to the choice of a particular model, which
are not included in Table~\ref{props.tab}, can then be quantified
and reduced by selecting the subset of models providing 
satisfactory simultaneous fits to both the optical and the near-IR data.

\subsection{Stochastic fluctuations}
\label{stochastic.sec}

\begin{table*}
\caption[]{Average numbers and flux contributions of red supergiant stars
in single age populations containing $10^6$\,M$_{\odot}$ of stars.
Is counted as a red supergiant any star with an initial mass above
7\,M$_{\odot}$ and an effective temperature below 4900\,K. The IMFs are
from Salpeter (1955) and Kroupa (1993). The 
lower mass cut-off is 0.1\,M$_{\odot}$ for the upper lines, 0.6 for lower
lines.}
\label{RSGnumbers.tab}
\begin{tabular}{lcccccccc} \hline \hline
Age  & Absolute & Fractional & Fractional & Fractional &
         Absolute & Fractional & Fractional    & Fractional \\
(Myr) &
   number     & number   &  contribution & contribution &
   number     & number   & contribution  & contribution\\
  & (Salpeter) & (Salpeter) & at 1\,$\mu$m & at 2.2\,$\mu$m &
    (Kroupa) & (Kroupa) & at 1\,$\mu$m  & at 2.2\,$\mu$m \\
  &           &            & (Salpeter) & (Salpeter ) &
             &            & (Kroupa)   & (Kroupa) \\
\hline 
8 &  60   & 2.0\,10$^{-3}$  &  28\,\% & 44\,\% &
     40   & 1.8\,10$^{-3}$  &  27\,\% & 43\,\% \\
10 & 150  & 5.2\,10$^{-3}$  &  72\,\% & 87\,\% &
     100  & 4.9\,10$^{-3}$  &  68\,\% & 85\,\% \\
30 & 260  & 8.3\,10$^{-3}$  &  64\,\% & 86\,\% &
     220  & 1.0\,10$^{-2}$  &  59\,\% & 84\,\% \\
60 & 360  & 1.3\,10$^{-2}$  &  49\,\% & 73\,\% &  
     340  & 1.7\,10$^{-2}$  &  46\,\% & 70\,\% \\
\hline
 8 & 120 & 2.4\,10$^{-2}$ & 28\,\% & 44\,\%  & 
      60 & 9.1\,10$^{-3}$ & 27\,\% & 43\,\%  \\
 60 & 730 & 1.5\,10$^{-1}$ & 49\,\% & 73\,\% & 
      570 & 8.6\,10$^{-2}$ & 46\,\% & 71\,\% \\
\hline
\end{tabular}
\end{table*}

The issue of stochastic fluctuations is the result of the 
contrast between the small fractional number of bright stars in a stellar 
population and their large contribution to the 
luminosity (Girardi \& Bica 1993, 
Santos \& Frogel 1997,
Lan\c{c}on \& Mouhcine 2000\footnote{On p. 35 in that article, divide
by $l$ instead of $L$ in the second expression given for $\sigma_L/L$.}, 
Cervi\~no \& Luridiana 2006). Table~\ref{RSGnumbers.tab} 
lists the {\em average} relative and absolute numbers of red 
supergiant stars in synthetic single-age populations of total 
mass $10^6$\,M$_{\odot}$, as well as the contribution of these stars
to the flux at 1\,$\mu$m and 2.2\,$\mu$m.
Note that these numbers depend on the shape of the IMF. 
Poisson statistics tell us that the
r.m.s dispersion in the actual number of red supergiants between
single clusters of that mass equals the square root of the
average number. In the K band, the red supergiants are
so much more luminous than other coeval stars that these variations
translate almost directly into dispersions in the fluxes
(Lan\c{c}on \& Mouhcine 2000). For colours such as V-K or Z-K, fluctuations are
also due partly to varying numbers of luminous blue stars. When
average numbers are small ($<10$), the predictions can be offset 
from the mean or  multimodal: 
the colours vary wildly depending on whether or not a
handful of massive stars happen to be in a red or in a blue
phase of their evolution. 

Table~\ref{RSGnumbers.tab} tells us it
is necessary to work with clusters of $\sim 10^6$\,M$_{\odot}$ or
more, if one wishes to test detailed model ingredients (unless large
samples of clusters are available for a statistical analysis).  
For instance, at an age of 30\,Myr a cluster of $\sim 10^6$\,M$_{\odot}$
will contain $260\pm 16$ red supergiants, that will on average
provide 85\,$\%$ of the K band flux. Clearly, the $\sim 6\,\%$ 
fluctuations in the red supergiant numbers (combined with fluctuations
in the numbers of blue stars) will produce a spread in colours
such as V-K that make it impossible to test the small differences 
in average properties resulting from conservative changes in the IMF.

The differences between evolutionary tracks are much larger than those
associated with changes in the IMF. We expect that at least some
of these will be testable with individual clusters more massive
than a few $10^5$\,M$_{\odot}$. Detailed computations with the
various sets of tracks now available will be needed 
in order to verify this statement.
As a general rule, even with masses around 10$^6$\,M$_{\odot}$,
the existence of the fundamental stochastic limitation must be
kept in mind.

\section{Conclusion}
\label{conclusion.sec}

Using new synthetic spectra of single stellar populations that extend
to 2.5\,$\mu$m and include red supergiant spectra at resolution 
$\lambda/\delta \lambda \simeq 750$,
we have analysed the near-IR spectra of a few of the most massive
star clusters in the starburst galaxy M\,82. 

We demonstrate that very good fits to all the near-IR photospheric
features seen at this resolution can now be obtained. In particular, 
the new synthetic spectra can be used for a precise subtraction
of the stellar background in emission line studies. The models also
significantly improve predictions of the low resolution energy
distribution around 1\,$\mu$m over those based on previous stellar 
libraries. Further improvements around 1\,$\mu$m are to be
expected from the inclusion of stars warmer than 5000\,K in the 
library of near-IR spectra used here.

Between ages of 10 and 60\,Myr, we have found that the degeneracy
between age and extinction is essentially perfect in near-IR 
broad band photometry (Z,Y,J,H,K), 
when the extinction law of Cardelli et al. (1989) is used.
A simultaneous fit of the near-IR spectra {\em and} the
optical\,+\,near-IR energy distribution was obtained 
for cluster M82-L, and this required a modified extinction law 
(e.g. the law of Cardelli et al., 1989, with R$_V<3.1$). This 
new evidence for non-standard extinction laws towards star 
clusters in starburst galaxies, which is not surprising considering
the very inhomogeneous spatial distribution of the dust in these
objects, is an additional difficulty in any attempt to derive
ages from photometry alone. The use of extended spectra allows to
constrain both the ages and the shape of the extinction law.
It is worth recalling, however, that a given shape of the extinction
law can correspond to a variety of values of the total amount of 
obscuration (e.g. Witt \& Gordon 2000). 

A table of red supergiant numbers and flux contributions has been
provided. We argue that the stochastic nature of the stellar mass
function must be kept in mind at all cluster masses, but that
with masses above 10$^5$\,M$_{\odot}$ the analysis of
individual extended spectra nevertheless allows us to test
{\em selected} parameters of the population synthesis models.
For instance, we expect it will be possible to exploit the largest
differences between various sets of current stellar evolution
tracks (e.g. tracks for rotating and for non-rotationg stars). But
we discourage attempts to test the shape of the upper IMF unless
cluster mass approaches 10$^6$\,M$_{\odot}$.

The absolute ages derived from the near-IR spectra 
depend on model parameters that are still highly uncertain, to 
a large part because the physics of red supergiants (evolution,
spectra) are particularly complex.
More work on the stellar models and more confrontations with
star cluster data will be needed. Care needs to be taken in 
matching the apertures of multi-wavelengths observations.
It is promising that data
quality now allows us to exploit ``details" that were neglected
until now, such as the differences between
spectra of supergiants of class Ia, Iab and Ib/II. 

The near-IR ages found with the current model assumptions for the 
observed IR-bright clusters in M\,82 are concentrated between 9 and 35\,Myr.
Indeed, their spectra display deep bands of CN and CO, and therefore favour 
the model ages at which the contributions of luminous red supergiants are
strongest. Cluster F, with weaker bands, is the oldest cluster of 
our sample. In most cases, the near-IR molecular bands of the models
at the ages derived from {\em optical} studies are marginally acceptable
or too weak. Changing the adopted evolutionary tracks or the 
parameters assigned to the spectra of the input stellar library can
result in deeper near-IR bands over a wider range of ages. Work is 
in progress to perform these tests, which should in time allow us to select 
the theoretical model inputs most appropriate for the 
analysis of star clusters in starburst galaxies.

\begin{acknowledgements}
The authors thank P.R. Wood, P. Hauschildt, W.D. Vacca, 
N. F\"orster Schreiber and R. O'Connell for fruitful discussions
in the preparation of this work.
\end{acknowledgements}



\begin{thebibliography}{} 
\bibitem{AlonsoHer03} Alonso-Herrero, A., Rieke, G.H., Rieke, M.J. \&
    Kelly, D.M. 2003, AJ 125, 1210
\bibitem{AndersGrev89} Anders, E., \& Grevesse, N. 1989, 
 Geochim. Cosmochim. Acta 53, 197
\bibitem{Aspletal05} Asplund, M., Grevesse, N., \& Sauval, A.J. 2006, in
  Cosmic Abundances as Records of Stellar Evolution and Nucleosynthesis,
  T.G. Barnes, III, \& F.N. Bash (eds.), ASP Conf. Ser. 336, 25
\bibitem[Bastian et al. (2007)]{Bastian07}
     Bastian, N., Konstantopoulos, I.,  Smith, L.J. et al.
     2007, MNRAS, 379, 1333
\bibitem{Bressanetal93} Bressan, A., Fagotto, F., Bertelli, G., \&
   Choisi, C. 1993, A\&AS 100, 647
\bibitem[Bruzual \& Charlot (2003)]{BruzChar03}
     Bruzual, G. \& Charlot, S. 2003, MNRAS 344, 1000
\bibitem[Cardelli et al. (1989)]{Cardelli89}
     Cardelli, J.A., Clayton, G.C. \& Mathis, J.S. 1989,
     ApJ  345, 245
\bibitem[Cenarro et al. (2001)]{CenarroLib01}
    Cenarro, A.J., Cardiel, N., Gorgas, J. et al.
    2001, MNRAS 326, 959
\bibitem[Cervi\~no \& Luridiana (2006)]{CervLuri06}
     Cervi\~no, M., \& Luridiana, V. 2006, A\&A 451, 475
\bibitem{Cushing04} Cushing, M.C., Vacca, W.D., \& Rayner, J.T. 2004,
     PASP 116, 362
\bibitem{deGrijsetal01} de Grijs, R., O'Connell, W.O. \& Gallagher III, J.S.
  2001, AJ 121, 768
\bibitem[Elmegreen (2006)]{Elmeg06}
      Elmegreen, B.G. 2006, ApJ 648, 572
\bibitem[Fioc \& Rocca-Volmerange (1997)]{FRV97}
     Fioc, M. \& Rocca-Volmerange, B. 1997, A\&A 326, 950
\bibitem[Fioc \& Rocca-Volmerange (1999)]{FRV99}
     Fioc, M., \& Rocca-Volmerange, B. 1999, astro-ph/9912179
\bibitem[F\"orster Schreiber et al. 2001]{FSetal01}
     F\"orster Schreiber, N.M., Genzel, R., Lutz, D., Kunze, D., \& 
     Sternberg, A.  2001, ApJ 552, 544
\bibitem[Gallagher \& Smith (1999)]{GalSmith99}
     Gallagher, J.S. \& Smith, L.J. 1999,
     MNRAS 304, 540
\bibitem[Girardi \& Bica (1993)]{GirBica93}
     Girardi, L., \& Bica, E. 1993, A\&A 274, 279
\bibitem[Gonz\'alez Delgado et al. (2005)]{GDetal05}
     Gonz\'alez Delgado, R.M., Cervi\~no, M., Martins, L.P., 
     Leitherer, C., \& Hauschildt, P.H. 2005, MNRAS 357, 945
\bibitem[Gray et al. (2001)]{Gray01}
     Gray, R.O., Graham, P.W., \& Hoyt, S.R. 2001, AJ 121, 2159
\bibitem{IvanovLib04}
    Ivanov, V.D., Rieke, M.J., Engelbracht, C.W. et al.
    2004, ApJS 151, 387
\bibitem[Kroupa et al. (1993)]{Kroupa93}
     Kroupa, P., Tout, C.A., \& Gilmore, G. 1993, MNRAS 262, 545
\bibitem[Lan\c{c}on \& Rocca-Volmerange (1996)]{LRV96}
    Lan\c{c}on, A. \& Rocca-Volmerange, B. 1996, NewA 1, 215
\bibitem[Lan\c{c}on \& Wood (2000)]{LW2000}
     Lan\c{c}on, A. \& Wood, P.R. 2000,
     A\&AS 146, 217
\bibitem[Lan\c{c}on \& Mouhcine (2000)]{LanconMouh00}
     Lan\c{c}on, A., \& Mouhcine, M. 2000, in Massive Stellar Clusters,
     ed. A. Lan\c{c}on, \& C.M. Boily, ASP Conf. Ser. 211, 34
\bibitem[Lan\c{c}on \& Mouhcine (2002)]{LM02}
     Lan\c{c}on, A., \& Mouhcine, M. 2002, A\&A 393, 167
\bibitem[Lan\c{c}on et al. (2007\,a)]{LHLM07}
     Lan\c{c}on, A., Hauschildt, P., Ladjal, D. \& Mouhcine, M., 2007\,a,
     A\&A, 468, 205 (LHLM07)
\bibitem[Lan\c{c}on et al. (2007\,b)]{Lancon07_LaPalma}
     Lan\c{c}on, A., Gallagher, J.S., de Grijs, R. et al. 2007\,b,
     in Stellar Populations as Building Blocks of Galaxies,
     Vazdekis, A. \& Peletier, R.F., eds. (Cambridge University Press),
     IAUS 241, 152
\bibitem[Larson 2006]{Larson06}
     Larsen, S.S. 2006, in Planets to Cosmology: Essential Science in
	the Final Years of the HST, ed. M. Livio \& S. Casertano
        (Cambridge Univ. Press), STScI Symp. Series 18, 35 
\bibitem[Lejeune et al. (1998)]{Lejeune98}
     Lejeune, T., Cuisinier, F., \& Buser, R. 1998, A\&AS 130, 65
\bibitem{Levesqetal05} Levesque, E.M., Massey, P., Olsen, K.A.G. et al.
  2005, ApJ 628, 973
\bibitem{Levesquetal06} Levesque, E.M., Massey, P., Olsen, K.A.G. et al.
  2006, ApJ 645, 1102
\bibitem{MaedMeyn01VII} Maeder, A., \& Meynet, G. 2001, A\&A 373, 555
\bibitem{Maraston05} Maraston, C. 2005, MNRAS 362, 799
\bibitem[McCrady et al. (2003)]{McCrady03} McCrady, N., Gilbert, A.M.,
   \& Graham, J.R. 2003, ApJ 596, 240
\bibitem[McCrady et al. (2005)]{McCrady05}
     McCrady, N., Graham, J.R. \& Vacca, W.D. 2005,
     ApJ 621, 278
\bibitem[McCrady \& Graham (2007)]{McCradyGra07}
     McCrady, N., \& Graham, J.R. 2007, ApJ 663, 844
\bibitem{McGregorCASPIR94} McGregor, P., Hart, J., Downing, M., Hoadley, D.,
    \& Bloxham, G. 1994, in Infrared Astronomy with Arrays, The Next Generation,
    ed. I.S. McLean, Astroph. and Sp. Sc. Lib. 190, 299
\bibitem[Meurer et al. (1995)]{Meurer95}
    Meurer, G.R., Heckman, T.M., Leitherer, C. et al.
    1995, AJ 110, 2665
\bibitem{Meyeretal99} Meyer, M.R., Edwards, S., Hinkle, K.H., \& Strom, S.E.
    1999, ApJ 508, 397
\bibitem[Meynet et al. (1994)]{Meynetal94}
   Meynet, G., Maeder, A., Schaller, G., Schaerer, D., \& Charbonnel, C. 1994,
   A\&AS 103, 97
\bibitem[Meynet \& Maeder (2000)]{MeynMaed00_V}
   Meynet, G. \& Maeder, A. 2000, A\&A 361, 101
\bibitem[Meynet \& Maeder (2003)]{MeynMaed03_X}
   Meynet, G. \& Maeder, A. 2003, A\&A 404, 975
\bibitem{MouhLan02} Mouhcine, M., \& Lan\c{c}on, A. 2002, A\&A 393, 167
\bibitem{Mutchler07} Mutchler, M., Bond, H.E., Christian, C.A. et al. 2007,
   PASP 119, 1
\bibitem{O'ConnMang78} O'Connell, R.W., \& Mangano, J.J. 1978, ApJ 221, 62
\bibitem{Oliva_al95} Oliva, E., Origlia, L., Kotilainen, J.K., 
   \& Moorwood, A.F.M. 1995, A\&A 301, 55
\bibitem{Origliaetal93} Origlia, L., Moorwood, A.F.M., \& Oliva, E. 1993,
   A\&A 280, 536
\bibitem{Origliaetal97} Origlia, L., Ferraro, F.R., Fusi Pecci, F., \& Oliva, E.
  1997, A\&A 321, 859
\bibitem{Pickles98} Pickles, A.J. 1998, PASP 110, 863
\bibitem{RaynerSpex03} Rayner, J.T., Toomey, D.W., Onaka, P.M. et al.
  2003, PASP 115, 362
\bibitem[Rieke \& Lebofsky (1979)]{RiekeLeb79} Rieke, G.H. \& Lebofsky, M.J.
   1979, ARA\&A 17, 477
\bibitem[Rieke et al. (1993)]{Rieke_al93}
  Rieke, G.H., Loken, K., Rieke, M.J., \& Tamblyn, P. 1993, ApJ 412, 99
\bibitem[Rossa et al. (2007)]{Rossa07}
  Rossa, J., Laine, S., van der Marel, R.P. et al. 2007, AJ 134, 2124
\bibitem[Salpeter (1955)]{Salpeter55}
    Salpeter, E.E., 1955, ApJ 121, 161
\bibitem[Santos \& Frogel (1997)]{SantosFrog97}
    Santos, J.F.C., Jr. \& Frogel, J.A. 1997, ApJ 479, 764
\bibitem[Satyapal et al. (1997)]{Satyap97}
   Satyapal, S., Watson, D.M., Pipher, J.L. et al. 1997, ApJ 483, 148
\bibitem[Smith \& Gallagher (2001)]{SmithGal01}
     Smith, L. J. \& Gallagher, J. S. 2001,
     MNRAS 326, 1027
\bibitem[Smith et al. (2006)]{SmithM82stis_06}
   Smith, L.J., Westmoquette, M.S., Gallagher, J.S., III, et al. 
   2006, MNRAS 370, 513
\bibitem[Smith et al. (2007)]{SmithM82B_07}
  Smith, L.J., Bastian, N., Konstantopoulos, I.S. et al. 2007,
  ApJ 667, L145
\bibitem{Tsuji76} Tsuji, T. 1976, PASJ 28, 543
\bibitem{VaccaSpeXA0V03} Vacca, W.D., Cushing, M.C., \& Rayner, J.T. 2003,
  PASP 115, 389
\bibitem[V\'azquez et al. (2007)]{Vazqetal07}
   V\'azquez, G.A., Leitherer, C., Schaerer, D., Meynet, G., \& Maeder, A. 2007
   ApJ 663, 995
\bibitem{Wallaceetal00} Wallace, L., Meyer, M.R., Hinkle, K.H., \& Strom, S.E.
  2000, ApJ 535, 325
\bibitem{WittGordon00} Witt, A.N., \& Gordon, K.D. 2000, ApJ 528, 799



\end{thebibliography}
\end{document}